%% file: main.tex
\documentclass[12pt,a4paper]{scrartcl}

\pdfoutput=1

\usepackage[utf8]{inputenc}
\usepackage[T1]{fontenc}
\usepackage{amsmath,amssymb,mathtools}
\usepackage{booktabs,tabularx, multirow}
\usepackage{graphicx,subfig}
\usepackage{xspace}
\usepackage[usenames]{xcolor}\definecolor{fscolor}{RGB}{44,118,255}
\usepackage{listings}
\usepackage[absolute]{textpos}
\usepackage[many]{tcolorbox}
\usepackage{xparse}
\usepackage[noblocks]{authblk}
\usepackage[font=small,labelfont=bf,format=plain,margin=0.05\textwidth]{caption}
\usepackage{bbm}
\usepackage{tabularx}
\usepackage{soul}
\usepackage{hyperref}
\usepackage[capitalize]{cleveref}
\usepackage{lineno}
\usepackage{lmodern}
\usepackage{microtype}

\usepackage[sorting=none,%
  citestyle=numeric-comp,%
  bibstyle=mynumeric,%
  giveninits=true]{biblatex}

\allowdisplaybreaks

\input{abbr.tex}

\addtokomafont{disposition}{\rmfamily}

\title{\Large Constraining \cp-violation in the Higgs--top-quark interaction using machine-learning-based inference}
\author[1,2]{Henning Bahl\footnote{\href{mailto:henning.bahl@desy.de}{hbahl@uchicago.edu}}}
\affil[1]{University of Chicago, Department of Physics, 5720 South Ellis Avenue, Chicago, IL~60637~USA}
\author[2]{Simon Brass\footnote{\href{mailto:simon.brass@desy.de}{simon.brass@desy.de}}}
\affil[2]{Deutsches Elektronen-Synchrotron DESY,
       Notkestraße~85,
       22607 Hamburg, Germany}
\date{}
\titlehead{\hfill \parbox{3cm}{\vspace{-2cm} \begin{flushright}\texttt{DESY 21-160} \texttt{EFI 21-6}\end{flushright}}}

\begin{document}
\maketitle

\begin{abstract}\noindent

\input{abstract.tex}

\end{abstract}

\newpage
\def\thefootnote{\arabic{footnote}}


\section{Introduction}
\label{sec:intro}

\input{sec_intro.tex}


\section{Effective model}
\label{sec:model}

\input{sec_model.tex}


\section{Methodology}
\label{sec:methods}

\input{sec_methods.tex}


\section{Event simulation and selection}
\label{sec:event_simulation}

\input{sec_MC.tex}


\section{Results}
\label{sec:results}

\input{sec_results.tex}


\section{Conclusions}
\label{sec:conclusions}

\input{sec_conclusions.tex}


\section*{Note added}

During the final stage of this project, \ccite{Barman:2021yfh} appeared following a similar approach to constrain \cp violation in the top-Yukawa interaction.


\section*{Acknowledgments}
\sloppy{
We thank Tim Stefaniak for collaboration in an early stage of the project. We acknowledge support by the Deutsche Forschungsgemeinschaft (DFG, German Research Foundation) under Germany's Excellence Strategy -- EXC 2121 ``Quantum Universe'' – 390833306. HB acknowledges support by the Alexander von Humboldt foundation.
}


\appendix





\clearpage
\printbibliography

\end{document}

%% file: abbr.tex
\newcommand{\cp}{\ensuremath{{\cal CP}}\xspace}
\newcommand{\SM}{{\text{SM}}}

\newcommand{\gev}{\,\, \mathrm{GeV}}

\newcommand{\invfb}{\,\, \mathrm{fb}^{-1}}

\definecolor{Darkgreen}{rgb}{0.,.7,0.2}
\definecolor{Darkblue}{rgb}{0.,.2,0.7}

\newcommand{\cv}{\ensuremath{c_V}\xspace}
\newcommand{\ct}{\ensuremath{c_t}\xspace}
\newcommand{\cttilde}{\ensuremath{\tilde c_{t}}\xspace}

\NewBibliographyString{refname}
\NewBibliographyString{refsname}
\DefineBibliographyStrings{english}{%
  refname = {Ref\adddot},
  refsname = {Refs\adddot}
}
\DeclareCiteCommand{\ccite}
  {%
  \ifnum\thecitetotal=1
    \bibstring{refname}%
  \else%
    \bibstring{refsname}%
  \fi%
  \addnbspace\bibopenbracket%
  \usebibmacro{cite:init}%
   \usebibmacro{prenote}}
  {\usebibmacro{citeindex}%
   \usebibmacro{cite:comp}}
  {}
  {\usebibmacro{cite:dump}%
   \usebibmacro{postnote}%
   \bibclosebracket}
\newrobustcmd*{\Ccite}{\bibsentence\ccite}

%% file: abstract.tex
While \cp violation in the Higgs interactions with massive vector boson is already tightly constrained, the \cp nature of the Higgs interactions with fermions is far less constrained. In this work, we assess the potential of machine-learning-based inference methods to constrain \cp violation in the Higgs top-Yukawa coupling. This approach enables the use of the full available kinematic information. Concentrating on top-associated Higgs production with the Higgs decaying to two photons, we derive expected exclusion bounds for the LHC and the high-luminosity phase of the LHC. We also study the dependence of these bounds on the Higgs interaction with massive vector bosons and their robustness against theoretical uncertainties. In addition to deriving expected exclusion bounds, we discuss at which level a non-zero \cp-violating top-Yukawa coupling can be distinguished from the SM. Moreover, we analyze which kinematic distributions are most sensitive to a \cp-violating top-Yukawa coupling.

%% file: sec_intro.tex
After the discovery of a particle consistent with the Standard Model (SM) Higgs boson --- within the current theoretical and experimental uncertainties --- by the ATLAS and CMS collaborations at the LHC~\cite{Aad:2012tfa,Chatrchyan:2012xdj}, the determination of its couplings and quantum numbers is one of the most important task for future LHC runs (as well as future colliders). One especially interesting property is the \cp nature of the Higgs boson interactions. While the amount of \cp violation in the SM is not sufficient to explain the baryon asymmetry of the Universe~\cite{Gavela:1993ts,Huet:1994jb}, the interactions of the Higgs boson can provide additional sources of \cp violation in many beyond SM (BSM) theories.

\cp violation in the Higgs sector can be constrained in two different ways: In the direct approach, \cp-odd observables are measured. For collider studies, these are typically constructed using angular distributions (e.g.\ for the Higgs decay into two tau leptons~\cite{CMS:2020rpr,CMS:2021sdq}). Measuring a non-zero value for a \cp-odd observable is an unambiguous sign of \cp violation. Therefore, such measurements, which are often challenging experimentally, are typically comparably model independent. As an alternative approach, the indirect approach can provide important complementary information. In the indirect approach, the effect of \cp-odd interactions on \cp-even observables is considered (e.g.\ the effect of a \cp-odd top-Yukawa coupling on the total top-associated Higgs production cross section). This indirect approach can yield strong constraints on \cp-violating interactions. It is, however, not guaranteed that deviations from the SM originate from \cp-violating interactions but they can in principle also be caused by \cp-even interactions. Consequently, these type of constraints are often associated with a larger model dependence.

Electric dipole moments (EDMs) are prominent examples for \cp-odd observables.
While recent EDM measurements (see e.g.~\ccite{Andreev:2018ayy,Abel:2020gbr}) severely constrain \cp violation in the Higgs sector~\cite{Brod:2013cka,deVries:2017ncy,deVries:2018tgs,Fuchs:2020uoc}, different contributions to the electric dipole moments --- originating e.g.\ from \cp violation in more than one Higgs coupling --- can cancel each other~\cite{Chien:2015xha,Cirigliano:2016nyn,Panico:2018hal}. Therefore, searching for \cp violation in the Higgs sector at colliders is an important complementary approach, especially since collider studies allow to disentangle the different couplings.

LHC studies have already ruled out the hypothesis of a pure \cp-odd Higgs boson~\cite{Khachatryan:2014kca,Aad:2015mxa}. A \cp-mixed Higgs boson is, however, far less constrained. Most existing experimental studies concentrate on the Higgs interaction with massive vector bosons~\cite{Aad:2015mxa,Aad:2016nal,Sirunyan:2017tqd,Sirunyan:2019nbs,Sirunyan:2019twz,Aad:2020mnm,CMS:2020dkv} (investigating Higgs production via vector-boson fusion and/or the Higgs decay into $W$ or $Z$ bosons). \cp violation in the Higgs interaction with massive vector boson can, however, only be induced at the loop level in a generic BSM theory and is, therefore, expected to be small. In contrast, \cp violation in the Higgs interaction with fermions can occur unsuppressed making an investigation of the Higgs--fermion--fermion interactions especially interesting from a phenomenological point of view. Existing experimental studies target \cp violation in the Higgs coupling to tau leptons~\cite{CMS:2020rpr,CMS:2021sdq} and the Higgs coupling to top quarks~\cite{Sirunyan:2020sum,Aad:2020ivc,CMS:2020dkv,CMS:2021nnc}. In addition, also \cp violation in the effective Higgs coupling to gluons has been constrained experimentally~\cite{ATLAS:2021pkb}.

In the present study, we focus on \cp violation in the Higgs interaction with top quarks, a possibility which has received considerable attention in the literature~\cite{Agrawal:2012ga,Ellis:2013yxa,Demartin:2014fia,Demartin:2015uha,Demartin:2016axk,Kobakhidze:2016mfx,Cao:2019ygh,Buckley:2015vsa,Gritsan:2016hjl,Azevedo:2017qiz,Barger:2018tqn,Goncalves:2018agy,Freitas:2012kw,Djouadi:2013qya,Boudjema:2015nda,Hou:2018uvr,Ren:2019xhp,Kraus:2019myc,Bahl:2020wee,Martini:2021uey}. While a \cp-violating top-Yukawa coupling affects many processes (e.g.\ Higgs production via gluon fusion, Higgs decay into two photons, $t\bar t$ production, etc.), top-associated Higgs production is the most direct probe since the top-Yukawa coupling appears at the tree level. Even though \cp-odd variables for top-associated Higgs production have been proposed~\cite{Mileo:2016mxg,Faroughy:2019ird,Bortolato:2020zcg,Goncalves:2021dcu}, the current experimental studies~\cite{Sirunyan:2020sum,Aad:2020ivc,CMS:2020dkv,CMS:2021nnc} --- targeting the Higgs decay to two photons --- exploit the effects of the \cp-odd Yukawa coupling on the kinematic distributions --- effectively mixing \cp-even and \cp-odd observables.

From a technical point of view, the studies presented in \ccite{Sirunyan:2020sum,Aad:2020ivc,CMS:2020dkv,CMS:2021nnc} rely on the use of boosted decision trees (BDTs) in order to separate the signal process from background as well as to distinguish a \cp-conserving from a \cp-violating top-Yukawa interaction. While the use of BDTs is a well-established tool for collider analyses, this method relies on binning the events in a discrimination observable, which is optimized using the BDTs. This reduction of the high-dimensional data into a low-dimensional summary statistics is unavoidably associated with a loss of information.

Novel analysis techniques promise to use the full information available in the event data. The main goal of the present paper to evaluate the potential of machine-learning-based inference methods to probe a \cp-odd top-Yukawa coupling. This approach was developed in \ccite{Brehmer:2019bvj,Brehmer:2018hga,Brehmer:2018kdj,Brehmer:2018eca,Stoye:2018ovl}. It uses machine learning to approximate the full likelihood fully taking into account parton shower as well as detector effects, which are only approximated in similar approaches like the matrix element method (see e.g.~\ccite{D0:2004rvt,Gao:2010qx,Alwall:2010cq,Bolognesi:2012mm,Avery:2012um,Andersen:2012kn,Artoisenet:2013vfa,Campbell:2013hz,Gainer:2013iya,Schouten:2014yza,Martini:2015fsa,Gritsan:2016hjl,Martini:2017ydu,Kraus:2019qoq}) or the optimal observable approach~\cite{Chang:2014rfa,Yue:2014tya,He:2014xla}.

From a practical point of view, we use the implementation of machine-learning-based inference in the tool \texttt{Madminer}~\cite{Brehmer:2019xox}. Similar to \ccite{Sirunyan:2020sum,Aad:2020ivc,CMS:2020dkv}, we focus on the Higgs decay to photons. We, however, only consider events containing at least one lepton. While this choice reduces the expected number of events, it significantly simplifies our analysis. In principle, it is, however, straightforward to also take into account events with no lepton as well as other Higgs decay channels. In addition to a modified Higgs--top-quark interaction, we also allow for deviations of the Higgs interaction to massive vector bosons ($HVV$ coupling) with respect to the SM, which is the most relevant Higgs coupling in top-associated Higgs production besides the top-Yukawa coupling. This allows us to study the dependence of the top-Yukawa coupling constraints on the $HVV$ coupling. Moreover, we study the impact of theoretical uncertainties by treating the renormalization scale as a nuisance parameter.

Using this setup, we derive expected limits on a \cp-violating Yukawa coupling using the currently available luminosity. In addition, we derive projections using the full LHC and High-Luminosity LHC (HL-LHC) data. In addition to assuming SM data, we also investigate for an exemplary case of a \cp-admixed Higgs boson what amount of data is needed to establish a deviation from the SM. Moreover, we evaluate which observables are most sensitive to the \cp character of the Higgs top-Yukawa coupling by evaluating the Fisher information.

Our paper is organized as follows. In \cref{sec:model}, we introduce the effective model used for our study. The employed methods are reviewed in \cref{sec:methods}. We give details on the event simulation and selection in \cref{sec:event_simulation}. The results are presented in \cref{sec:results}. We conclude in \cref{sec:conclusions}.

%% file: sec_model.tex
For the present study, we employ an effective Higgs model (based on the SM), which is closely related to the ``Higgs characterisation'' model defined in~\ccite{Artoisenet:2013puc,Maltoni:2013sma,Demartin:2014fia}.

We are most interested in the top-Yukawa part of the Lagrangian which is given by
\begin{align}\label{eq:topYuk_lagrangian}
\mathcal{L}_\text{top-yuk} = - \frac{y_t^\SM}{\sqrt{2}} \bar t \left(\ct + i \gamma_5 \cttilde\right) t H.
\end{align}
Here, $H$ is the Higgs field and $t$ is the top-quark field. The prefactor $y_t^\SM$ is the SM top-Yukwawa coupling. Deviations from the SM are parameterized in terms of \ct and \cttilde. \ct modifies the \cp-even part of the top-Yukawa interaction, whereas \cttilde induces a \cp-odd top-Yukawa interaction. The SM is recovered for $\ct = 1$ and $\cttilde = 0$. In an effective field theory framework, the deviations from SM can be thought of to be induced by dimension-6 operators (see e.g.~\ccite{Demartin:2015uha}). \ct and \cttilde correspond to $\kappa_t$ and $\tilde\kappa_t$ as used in~\ccite{Sirunyan:2020sum}, respectively.

Instead of the coupling modifiers \ct and \cttilde, also an absolute value, denoted by $|g_t|$ and a \cp-violating phase $\alpha$ are often used to parameterize the top-Yukawa interaction. They are related to \ct and \cttilde via
\begin{align}
|g_t| \equiv \sqrt{\ct^2 + \cttilde^2}, \qquad \qquad \tan\alpha = \frac{\cttilde}{\ct}.
\label{eq:alpha_gt}
\end{align}
The quantities $|g_t|$ and $\alpha$ correspond to $\kappa_t$ and $\alpha$ as used in~\ccite{Aad:2020ivc}.

Top-associated Higgs production --- the target of this study ---, however, depends not only on the Higgs--top-quark interaction but also on the Higgs interaction with massive vector bosons. We take into account a $SU(2)_L$ preserving modification of the SM interaction,
\begin{align}\label{eq:HVV}
\mathcal{L}_V = \cv H \left(\frac{M_Z^2}{v} Z_\mu Z^\mu + 2\frac{M_W^2}{v} W_\mu^+ W^{-\mu}\right),
\end{align}
where $Z$ and $W$ denote the massive vector boson fields and $M_{Z,W}$ their respective masses ($v\simeq 246$~GeV is the Higgs vacuum expectation value). The Higgs interaction with massive vector bosons is rescaled by the common factor \cv. In principle, it is also possible to include additional operators going beyond the form of the SM operators (e.g.\ operators of the form $H Z_{\mu\nu}Z^{\mu\nu}$ with $Z_{\mu\nu}$ being the $Z$ boson field strengths). Since the main focus of the present study is the Higgs--top-quark interaction, we omit these additional operators. Moreover, we do not expect that their presence would be distinguishable from a deviation in the SM-like $HVV$ couplings (see \cref{eq:HVV}) when investigating top-associated Higgs production --- the target process of the present study.

Note that the modified Higgs--top-quark interaction of \cref{eq:topYuk_lagrangian} also affects the Higgs decay into two photons, which will be our target decay process in the next Section. The modification of the Higgs decay rate to two photons can result in stringent constraints on the parameter space (see e.g.~\ccite{Bahl:2020wee}). In the present work, we will, however, assume the Higgs decay rate to two photons to be SM-like, which can e.g.\ be achieved by the presence of at least one electrically charged BSM particle decorrelating the Higgs decay to two photons from the Higgs--top-quark interaction.

%% file: sec_methods.tex
In this Section, we discuss the concept of machine-learning-based inference~\cite{Brehmer:2019bvj,Brehmer:2018hga,Brehmer:2018kdj,Brehmer:2018eca,Stoye:2018ovl} (as implemented in the tool \texttt{Madminer}). The discussion follows closely \ccite{Brehmer:2018kdj,Brehmer:2018eca,Brehmer:2019xox}, in which more details can be found.


\subsection{Estimating the likelihood}
\label{sec:likelihood}

When performing LHC measurements, a key object is the likelihood function $p_\text{full}(\{x_i\}|\theta)$ giving the probability of observing a set of events with the observables $x_i$ for a given model with parameters $\theta$. It can be written as
\begin{align}\label{eq:pfull}
p_\text{full}(\{x_i\}|\theta) = \text{Pois}(n|L\sigma(\theta))\prod_i p(x_i|\theta),
\end{align}
where $n$ is the number of events, $L$ the integrated luminosity, $\sigma(\theta)$ is the cross section as a function of the model parameters, and $\text{Pois}(n|\lambda) = \lambda^n e^{-\lambda}/n!$ is the probability mass function of the Poisson distribution. $p(x|\theta)$ denotes the probability density of observing a single event with observables $x$ for a given model with parameters $\theta$,
\begin{align}
p(x|\theta) = \frac{1}{\sigma(x)}\frac{d^d\sigma(x|\theta)}{dx^d},
\end{align}
where $d$ is the dimension of the vector $x$. We can sample the distribution $p(x|\theta)$ by using Monte-Carlo (MC) simulators. Typically, these work in three steps: First, parton-level events are generated according to the matrix element of the process; second, the parton-level events are processed by a parton shower accounting for the effects of soft radiation and hadronization; third, the detector response is simulated. These steps can be written symbolically in the form
\begin{align}\label{eq:pxtheta_int}
p(x|\theta) = \int dz_d \int dz_s \int dz_p \underbrace{p(x|z_d) p(z_d|z_s) p(z_s|z_p) p(z_p|\theta)}_{= p(x,z|\theta)}
\end{align}
with the latent variables $z_d$, $z_p$, and $z_d$. The variables $z_p$ describe the parton-level observables; $z_s$, the parton-shower history; and $z_d$, the response of the detector. We denote the integrand as $p(x,z|\theta)$. Due to the large number of latent variables, the integral cannot be computed. Therefore, while $p(x|\theta)$ can be sampled, it cannot be computed directly.

Traditionally, this problem is circumvented by restricing the analysis to a low-dimensional summary statistics (e.g.\ the invariant mass of a tentative resonance). The likelihood $p(x|\theta)$ can then be estimated using histograms. Other approaches, like the matrix element method (see e.g.~\ccite{Bolognesi:2012mm,Avery:2012um,Artoisenet:2013vfa,Gainer:2013iya,Gritsan:2016hjl}) or the optimal observable approach~\cite{Chang:2014rfa,Yue:2014tya,He:2014xla} try to approximate the integrals over $z_d$ and $z_p$ in \cref{eq:pxtheta_int} by suitable transfer functions.

Machine-learning-based inference avoids using low-dimensional summary statistics or approximating the effects of the parton shower or the detector. Instead, the likelihood or the likelihood ratio
\begin{align}
r(x|\theta_0,\theta_1) = \frac{p(x|\theta_0)}{p(x|\theta_1)},
\end{align}
where $\theta_0$ and $\theta_1$ represent two different parameter points, is estimated directly.

The estimation process can be improved by directly taking into account information from the MC simulator. This can be understood if looking at the joint likelihood ratio,
\begin{align}\label{eq:rxz}
r(x,z|\theta_0,\theta_1) &\equiv \frac{p(x,z|\theta_0)}{p(x,z|\theta_1)} = \nonumber\\
&= \frac{p(x|z_d)p(z_d|z_s)p(z_s|z_p)p(z_p|\theta_0)}{p(x|z_d)p(z_d|z_s)p(z_s|z_p)p(z_p|\theta_1)} = \nonumber\\
&= \frac{p(z_p|\theta_0)}{p(z_p|\theta_1)} = \nonumber\\
&= \frac{d\sigma(z_p|\theta_0)}{d\sigma(z_p|\theta_1)}\frac{\sigma(\theta_1)}{\sigma(\theta_0)},
\end{align}
where $d\sigma(z_p|\theta)$ are the parton-level event weights which can be calculated via
\begin{align}
d\sigma(z_p|\theta) = \frac{(2\pi)^4 f_1(x_1,Q^2)f_2(x_2,Q^2)}{8 x_1 x_2 s} |\mathcal{M}|^2(z_p|\theta)d\Phi(z_p).
\end{align}
Here, $f_1$ and $f_2$ are the parton-distribution functions (PDF) depending on the momentum fractions $x_{1,2}$ and the momentum transfer $Q$. $s$ is the square of the center-of-mass energy; $\mathcal{M}$, the matrix element of the considered process; and $\Phi$, the corresponding phase space. The parton-shower and detector effects cancel in \cref{eq:rxz}. Since we can calculate the parton-level event weights reasonably fast (using the ``morphing'' reweighting technique~\cite{ATL-PHYS-PUB-2015-047,Brehmer:2018eca,Brehmer:2019xox}), the joint likelihood ratio is accessible.

Analogously, one can obtain the joint score,
\begin{align}
t(x,z|\theta) &\equiv \nabla_\theta \log p(x,z|\theta) =\nonumber\\
&= \frac{p(x|z_d)p(z_d|z_s)p(z_s|z_p)\nabla_\theta p(z_p|\theta)}{p(x|z_d)p(z_d|z_s)p(z_s|z_p)p(z_p|\theta)} =\nonumber\\
&= \frac{\nabla_\theta d\sigma(z_p|\theta)}{d\sigma(z_p|\theta)} - \frac{\nabla_\theta \sigma(\theta)}{\sigma(\theta)},
\end{align}
which can be calculated by evaluating the derivatives of the parton-level event weights and the total cross section.

As shown in~\ccite{Brehmer:2018eca,Brehmer:2018hga}, the joint likelihood ratio (and the joint score) can be used to define suitable loss functions whose minimizing function is the true likelihood ratio $r(x|\theta_0,\theta_1)$ (or the true score $t(x|\theta) \equiv \nabla_\theta\log p(x|\theta)$). A simple example for such a loss function is
\begin{align}\label{eq:loss_function}
L[\hat r(x|\theta_0,\theta_1)] = \frac{1}{N} \sum_{(x_i,z_i)\sim p(x,z|\theta_1)} |r(x_i, z_i |\theta_0,\theta_1) - \hat r(x_i|\theta_0,\theta_1)|^2,
\end{align}
where $\hat r(x|\theta_0,\theta_1)$ is the estimator for $r(x|\theta_0,\theta_1)$. The sum runs of events sampled according to the probability distribution $p(x,z|\theta_1)$ (using MC simulators). An analogous loss function can be defined for the score.

The minimization procedure is perfectly suited to be tackled by machine learning. The estimator function can be expressed in terms of a neural network (NN). The NN can then be trained with the loss function given in \cref{eq:loss_function} using standard techniques. In order to reduce the uncertainty associated with the training of the NN, an ensemble of NNs with different random seeds can be trained~\cite{lakshminarayanan2017simple}. For the results presented in \cref{sec:results}, we found this averaging to be essential in order to smoothen out fluctuations of the individual networks.

Note that in practice more involved loss functions, featuring faster and more stable convergence, are used. For our numerical study, we will employ the ALICES~\cite{Stoye:2018ovl} and SALLY~\cite{Brehmer:2018hga} methods.

The ALICES loss functional is given by
\begin{align}
  \label{eq:alices}
  &\mathcal{L}_{\text{ALICES}} [\hat{s}(x|\theta_0, \theta_1)] = \nonumber\\
  &= - \frac{1}{N} \sum_{(x_i, z_i) \sim p(x_i, z_i)} \Bigg[ s(x_i, z_i|\theta_0, \theta_1) \log \hat{s}(x_i) + (1 - s(x_i, z_i| \theta_0, \theta_1)) \log (1 - \hat{s}(x_i)) \nonumber\\
  & \hspace{3.7cm} + \alpha (1 - y_i) \left| t(x_i, z_i|\theta_0, \theta_1) - \left. \nabla_{\theta} \log \left( \frac{1 - \hat{s}(x_i | \theta, \theta_1)}{\hat{s}(x_i|\theta, \theta_1)}\right) \right|_{\theta_0} \right|^2 \Bigg],
\end{align}
where the function $s$ is defined via
\begin{align}
\hat r(x|\theta_0,\theta_1) = \frac{1 - \hat s(x|\theta_0,\theta_1)}{\hat s(x|\theta_0,\theta_1)}.
\label{eq:r_vs_s}
\end{align}
The first term in the square bracket parametrizes the deviation of the joint likelihood score $\hat r$ from the true likelihood ratio $r$. The second term in the square bracket --- proportional to the hyperparameter $\alpha$ --- includes an additional mean-squared error loss for the joint score.

The simpler SALLY loss functional --- with the goal to estimate the score --- reads
\begin{align}
  \label{eq:sally}
  \mathcal{L}_{\text{SALLY}} [\hat{s}(x|\theta_0, \theta_1)]= - \frac{1}{N}  \sum_{(x_i, z_i) \sim p(x_i, z_i)} \left| t(x,z|\theta_0) - \hat t (x|\theta_0)\right|^2
\end{align}
and basically corresponds to the second line of \cref{eq:alices}.


\subsection{Fisher information}
\label{sec:fisher}

While the likelihood is crucial to derive exclusion limits, it is not the most useful quantity to quickly assess the sensitivity of a certain measurement. For the purpose of optimizing an analysis or for comparing the sensitivity of different processes, often the Fisher information matrix is more useful~\cite{Brehmer:2016nyr,Brehmer:2017lrt}. It is defined via
\begin{align}
I_{ij}(\theta) = \mathbb{E}\left[\frac{\partial\log p_{\text{full}}(\{x\}|\theta)}{\partial\theta_i}\frac{\partial\log p_\text{full}(\{x\}|\theta)}{\partial\theta_j}\bigg|_\theta\right],
\end{align}
where the symbol ``$\mathbb{E}$'' is used to denote the expectation value.

According to the Cramér-Rao bound~\cite{Rao:1945,Cramer:1946}, the minimal covariance of an estimator $\hat\theta$ is given by the inverse of the Fisher matrix,
\begin{align}
\text{cov}(\hat\theta|\theta)_{ij} \ge I_{ij}^{-1}(\theta),
\end{align}
where we assumed that our estimator is unbiased (i.e.\ that its expectation value $\bar\theta$ is equal to the true parameter point $\theta$). $\text{cov}(\hat\theta|\theta)_{ij} \equiv \mathbb{E}\left[(\hat\theta_i -\bar\theta)(\hat\theta_i - \bar\theta)|\theta)\right]$ is the covariance matrix. In other words, the size of the Fisher information is directly related to the reachable precision level of the respective parameter. For example, the measurement error for a parameter $\theta$ in the one-dimensional case is given by $\Delta\theta \ge 1/\sqrt{I(\theta)}$.

For our purposes, we can calculate the Fisher matrix via
\begin{align}
I_{ij}(\theta) \simeq \frac{L}{\sigma(\theta)} \frac{\partial\sigma(\theta)}{\partial\theta_i}\frac{\partial\sigma(\theta)}{\partial\theta_j} + \frac{L \sigma(\theta)}{n} \sum_{x\sim p(x|\theta)} t_i(x|\theta)t_j(x|\theta),
\end{align}
where the first term encodes the information in the total rate of a process and the second term the information contained in the kinematic distributions.

%% file: sec_MC.tex
The target of our analysis is top-associated Higgs production consisting out of the sub-channels $t\bar t H$, $tH$, and $tWH$ production. We focus on the $H\rightarrow\gamma\gamma$ channel due to its sharp peak structure around the measured Higgs mass in the invariant mass spectrum of the photons. This feature allows for a subtraction of the background event rates originating from non-Higgs processes, whose invariant mass spectrum typically falls smoothly with increasing invariant mass. We assume that the non-Higgs background is already subtracted by a fit to the $m_{\gamma\gamma}$ distribution in the experimental data.

\begin{figure}
    \centering
    \includegraphics[width=0.20\textwidth]{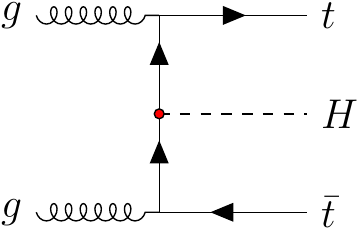}\hspace{.3cm}
    \includegraphics[width=0.25\textwidth]{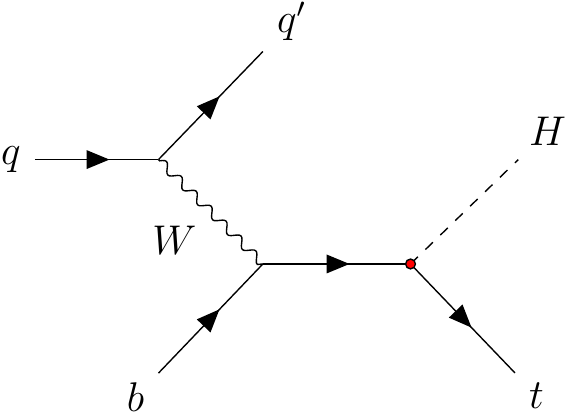}\hspace{.3cm}
    \includegraphics[width=0.15\textwidth]{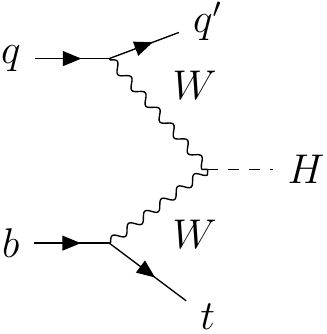}\\[.4cm]
    \includegraphics[width=0.20\textwidth]{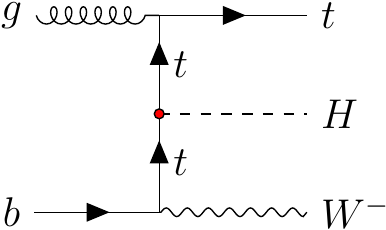}\hspace{.4cm}
    \includegraphics[width=0.25\textwidth]{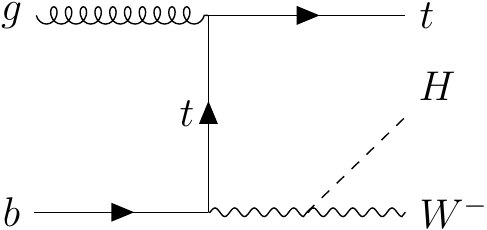}
    \caption{Exemplary Feynman diagrams for the $t\bar t H$ (upper left diagram), $tH$ (upper middle and right diagrams), and $tWH$ (lower diagrams) signal processes.}
    \label{fig:signal_feynman}
\end{figure}

\begin{figure}
    \centering
    \includegraphics[width=0.17\textwidth]{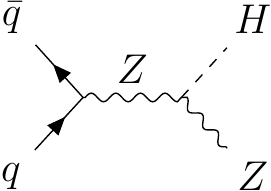}\hspace{1.5cm}
    \includegraphics[width=0.17\textwidth]{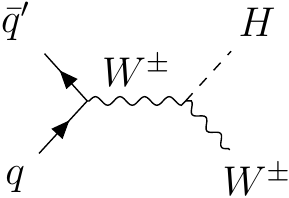}
    \caption{Exemplary Feynman diagrams for the $ZH$ and $WH$ background processes.}
    \label{fig:bkg_feynman}
\end{figure}

In order to simplify the analysis further, we also require at least one prompt lepton in the final state produced via a $Z$ or $W$ boson decay. Correspondingly, we consider $t\bar tH$, $tH$, and $tWH$ production as signal processes (see \cref{fig:signal_feynman}); $ZH$ and $WH$ production are considered as background processes (see \cref{fig:bkg_feynman}).\footnote{We neglect the $s$-channel contribution to $tH$ production as well as the gluon-induced $ZH$ production since their cross sections are small in comparison to the other involved processes.}

We use \texttt{MadGraph5\_aMC@NLO~2.8.2}~\cite{Alwall:2014hca} to generate MC event samples for these processes with \texttt{Pythia~8.244}~\cite{Sjostrand:2007gs} as parton shower (employing the A14 set of tuned parameters~\cite{ATL-PHYS-PUB-2014-021}). As parton-distribution function (PDF), we use the \texttt{MSTW2008LO}~\cite{Martin:2009iq} PDF set assessed via the \texttt{LHAPDF~6.2.3} interface~\cite{Whalley:2005nh}. The detector response is simulated using \texttt{Delphes~3.4.2}~\cite{deFavereau:2013fsa} employing either the ATLAS-LHC or the HL-LHC~\cite{Cepeda:2019klc,Atlas:2019qfx} configuration card provided by \texttt{Delphes}.\footnote{In the ATLAS-LHC card, we have modified the default setting for the radius parameter used by the anti-$k_t$ jet clustering algorithm to $R = 0.4$ instead of $R =0.6$. This choice is more often used in ATLAS analyses and also corresponds to the value used in the HL-LHC card.}  We assume the center-of-mass energy to be $\sqrt{s} = 13$ TeV for both the LHC and the HL-LHC.\footnote{The possible increase in the center-of-mass energy to $\sqrt{s} = 14$ TeV at the HL-LHC~\cite{Cepeda:2019klc,Atlas:2019qfx} would lead to an increase of the cross section for the top-associated Higgs production by about 20\%.}  We generate events for these processes at the leading-order (LO) level employing the ``Higgs characterization model''~\cite{Artoisenet:2013puc,Maltoni:2013sma,Demartin:2014fia}. The overall event rates for each subprocess are rescaled to state-of-the-art SM predictions~\cite{deFlorian:2016spz} using flat $K$ factors.

In order to account for theoretical uncertainties, we vary the renormalization and factorization scale in the interval $[1/2,2]$ times the central scale, which is the central squared transverse mass after $k_T$ clustering, using the built-in \texttt{MadGraph} reweighting functionality.\footnote{For top-associated Higgs production, the variation of the renormalization scale at LO results in a shift of the total cross section similar in size as the shift between the LO and NLO cross sections (see e.g.~\ccite{Demartin:2014fia,Demartin:2015uha,Demartin:2016axk}).} For the estimate of the likelihood, the renormalization scale is treated as an additional free parameter. After deriving the likelihood, we profile over the renormalization scale without preferring specific values. From a conceptional point of view, also the PDF uncertainty can be taken into account in exactly the same way. With the current version of \texttt{MadMiner} (version~\texttt{0.9}), we, however, found the profiling over the additional PDF nuisance parameters to significantly increase the computational cost to evaluate the final profiled likelihood. Therefore, we do not include this source of uncertainty in the present study. We expect that the run time can be reduced by further optimizations of \texttt{MadMiner}.

Using the setup described above, we generate MC event samples for 10 different benchmark points, which are distributed throughout our three-dimensional parameter space. One of these benchmark points represents the SM ($\cv = \ct = 1$ and $\cttilde = 0$). For the SM point, we generated $5\times 10^{5}$ events for each of the five different subprocesses ($t\bar tH$, $tH$, $tWH$, $ZH$ and $WH$); for the other benchmark points, we generated $1\times 10^{5}$ events per subprocess. Event weights for other parameter points, which are not identical to one of the benchmark points, are calculated using the morphing technique described in~\ccite{ATL-PHYS-PUB-2015-047,Brehmer:2018eca,Brehmer:2019xox}. As input for the neural network, a set of events is drawn from MC events with probabilities given by the respective event weights (a given MC event can appear multiple times). In a second step, these unweighted events are augmented by calculating the joint likelihood ratio $r(x,z)$ and the joint score $t(x,z)$ (see \cref{sec:methods}).

\begin{table}
\small
\begin{center}
\def\arraystretch{1.2}
\begin{tabular}{lc}
\toprule
observable & condition \\
\midrule
$N_{\gamma}$                                                             & $\ge 2$ (with $|\eta| < 2.5$ and $p_T > 25\gev$) \\
$(p_{T,1}^{\gamma}, p_{T,2}^{\gamma})$                                   & $\ge (35, 25)$ GeV                               \\
$m_{\gamma\gamma}$                                                       & $[105-160]$ GeV                                  \\
$(p_{T,1}^{\gamma}/m_{\gamma\gamma}, p_{T,2}^{\gamma}/m_{\gamma\gamma})$ & $\ge (0.35,0.25)$                                \\
$N_{\ell}$                                                               & $\ge 1$ (with $|\eta| < 2.5$ and $p_T > 15\gev$) \\
$m_{\ell \ell}$                                                          & $[80,100]$ GeV vetoed if same flavour            \\
$N_{jet}$                                                                & $\ge 1$ (with $|\eta| < 2.5$ and $p_T > 25\gev$) \\
\bottomrule
\end{tabular}
\end{center}
\caption{Summary of preselection cuts.}
\label{tab:preselection}
\end{table}

Following closely the experimental analyses, we impose preselection cuts on our event samples, which correspond to a simplified version of the typical preselection cuts as used by ATLAS and CMS for $t\bar tH, H\rightarrow\gamma\gamma$ measurements~\cite{Sirunyan:2020sum,Aad:2020ivc}. They are summarized in \cref{tab:preselection}. A very similar preselection was used in~\ccite{Bahl:2020wee}.

We require all events to feature at least two photons with a pseudo-rapidity $|\eta| < 2.5$ and $p_T > 25\gev$ fulfilling the default isolation criteria of the ATLAS and HL-LHC \texttt{Delphes} cards. The two highest-$p_T$ photons form the Higgs boson candidate. The highest-$p_T$ photon is required to have a transverse momentum, $p_{T,1}^{\gamma}$, larger than 35~GeV; the second-highest-$p_T$ photon, to have a transverse momentum, $p_{T,2}^{\gamma}$, larger than 25~GeV. The invariant mass of the di-photon system must be close to the Higgs boson mass (i.e.\ within in the range of $[105-160]$ GeV). Moreover, we impose that $p_{T,1}^{\gamma}/m_{\gamma\gamma} \ge 0.35$ and $p_{T,2}^{\gamma}/m_{\gamma\gamma} \ge 0.35$ in order to suppress photon radiation from initial-state partons.

As mentioned above, we also require at least one lepton with $p_{T,l} > 15$ GeV and $|\eta_l| < 2.5$ in the final state. For an event containing two leptons, the invariant mass of leptons must not lie within in the range $[80,100]$ GeV if the leptons have opposite charge and the same flavour in order to suppress $Z$-boson induced background (i.e.\ from $ZH$ production). Besides, we also require at least one jet with a $p_T > 20$ GeV and $|\eta_j| < 2.5$ in the final state.

After event generation and imposing the cuts listed above, a set of observables is calculated for each remaining event. This observable set is then used as input for the NN (see \cref{sec:methods}). The observable set includes the number of leptons ($n_l$), the number of photons ($n_\gamma$), the number of jets ($n_j$), the number of $b$ jets ($n_b$), the missing transverse energy ($E_{T,\text{miss}}$), the azimuth angle of the missing transverse energy vector ($\phi_{E_{T,\text{miss}}}$), the visible energy ($E_\text{vis}$), the pseudorapidity of the visible energy vector ($\eta_{E_\text{vis}}$), the transverse momenta of the two leading jets ($p_{T,j_1}$ and $p_{T,j_2}$), the azimuth angles of the two leading jets ($\phi_{j_1}$ and $\phi_{j_2}$), the pseudorapidities of the two leading jets ($\eta_{j_1}$ and $\eta_{j_2}$), the transverse momenta of the two leading photons ($p_{T,\gamma_1}$ and $p_{T,\gamma_2}$), the azimuth angles of the two leading photons ($\phi_{\gamma_1}$ and $\phi_{\gamma_2}$), the pseudorapidities of the two leading photons ($\eta_{\gamma_1}$ and $\eta_{\gamma_2}$), the transverse momenta of the two leading leptons ($p_{T,\ell_1}$ and $p_{T,\ell_2}$), the azimuth angles of the two leading leptons ($\phi_{\ell_1}$ and $\phi_{\ell_2}$), the pseudorapidities of the two leading leptons ($\eta_{\ell_1}$ and $\eta_{\ell_2}$), and the charges of the two leading leptons ($Q_{\ell_1}$ and $Q_{\ell_2}$). In addition to these low-level inputs, we also include several high-level observables which are computed out of the low-level observables: the rapidity difference of the two leading jets ($\Delta\eta_{j_1j_2}$), the azimuth angle difference of the two leading jets ($\Delta\phi_{j_1j_2}$), the invariant mass of the two leading jets ($m_{jj}$), the rapidity difference of the leading $b$ jet and the leading non-$b$ jet ($y_{bj}$), the invariant mass of the two leading photons ($m_{\gamma\gamma}$), the transverse momenta of the two leading photon system ($p_{T,H}$), the invariant mass of the two leading leptons ($m_{\ell\ell}$), and the transverse mass of leading $b$ jet and leading lepton system ($m_T^\text{top}$). We then use these 36~observables as input for the NN.

Some of these observables are only very weakly sensitive to the \cp character of the top-Yukawa coupling. We, nevertheless, take these observables into account in order to provide the NN with as much information as possible. The invariant masses of the two-leading photons, $m_{\gamma\gamma}$, as well as of the two leading leptons, $m_{\ell\ell}$ are included in order to impose the cuts summarized in \cref{tab:preselection}. 

\begin{figure}
    \centering
    \includegraphics[width=.95\textwidth]{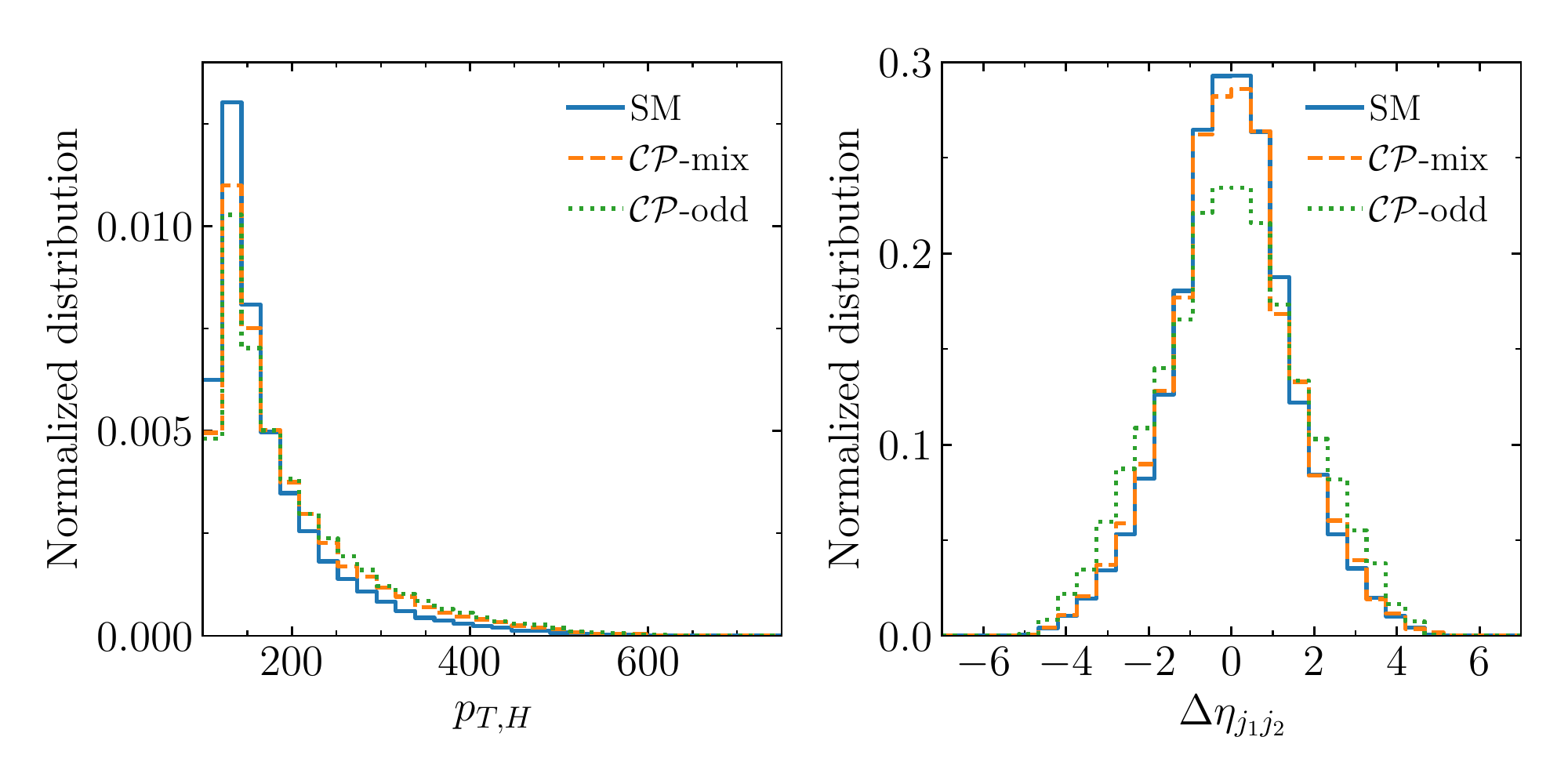}
    \caption{Normalized distribution of $p_{T,H}$ and $\Delta\eta_{j_1j_2}$ for three different parameter points: SM with $\ct = 1$, $\cttilde = 0$ and $\cv = 1$ (blue solid), \cp-mix with $\ct = \cttilde = 1.2$ and $\cv = 1$ (orange dashed), as well as \cp-odd with $\ct = 0, \cttilde = \cv = 1$ (green dotted).}
    \label{fig:distributions}
\end{figure}

As examples for observables strongly sensitive to the \cp nature of the top-Yukawa coupling (see also e.g.~\ccite{Demartin:2014fia}), we show in \cref{fig:distributions} the normalized distributions of the Higgs transverse momentum, $p_{T,H}$, and the pseudorapidity difference of the two leading jets, $\Delta\eta_{j_1j2}$ (events with only one jet are not shown in the histogram). The distributions are shown for three different parameter points: SM with $\ct = 1$, $\cttilde = 0$ and $\cv = 1$ (blue solid), \cp-mix with $\ct = \cttilde = 1.2$ and $\cv = 1$ (orange dashed), as well as \cp-odd with $\ct = 0, \cttilde = \cv = 1$ (green dotted). For the left panel of \cref{fig:distributions}, showing the distribution of $p_{T,H}$, we observe that the Higgs transverse momentum is on average harder in presence of a non-zero \cp-odd top-Yukawa coupling. For the right panel of \cref{fig:distributions}, showing the distribution of $\Delta\eta_{j_1j_2}$, we notice that for a \cp-odd top-Yukawa coupling the two leading jets tend to have a larger angular separation. We perform a more detailed analysis regarding the sensitivity of the different observables to the \cp character of the top-Yukawa coupling in \cref{sec:resuls_obs}.

%% file: sec_results.tex
In this Section, we present the tentative constraints on \ct and \cttilde using the methodology described in \cref{sec:methods}.

The limits shown in \cref{fig:expected_ct_cttilde,fig:expected_ct_cttilde_profile,fig:observed_ct_cttilde} are derived using the ALICES method as described in \cref{sec:likelihood}. For the LHC results, the likelihood is constructed using four different NNs, with two hidden layers, trained with a sample of $10^6$ unweighted events, over which we average by forming an ensemble~\cite{Brehmer:2018eca}; for the HL-LHC results, which cover a smaller parameter region, we use samples of $10^5$ unweighted events and average over six different neural networks.


\subsection{Expected limits at the LHC and the HL-LHC}

First, we look at the limits on \ct and \cttilde expected for SM data at the LHC using $139\invfb$ (corresponding roughly to the currently recorded event data) and $300\invfb$ (corresponding to event data recorded after LHC Run~3) as well the limits expected for SM data at the HL-LHC using $3000\invfb$.

\begin{figure}
    \centering
    \includegraphics[width=0.49\textwidth]{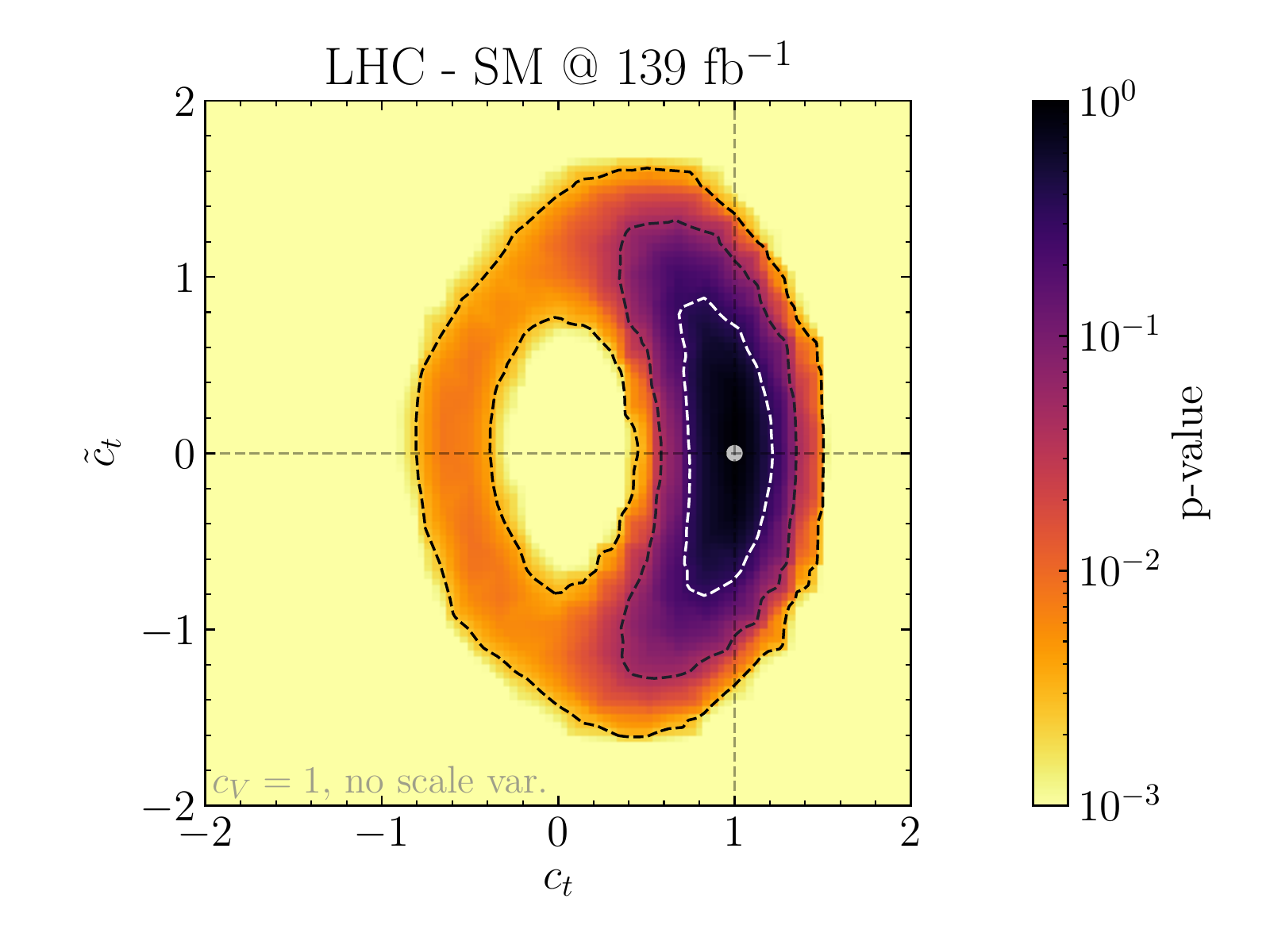}
    \includegraphics[width=0.49\textwidth]{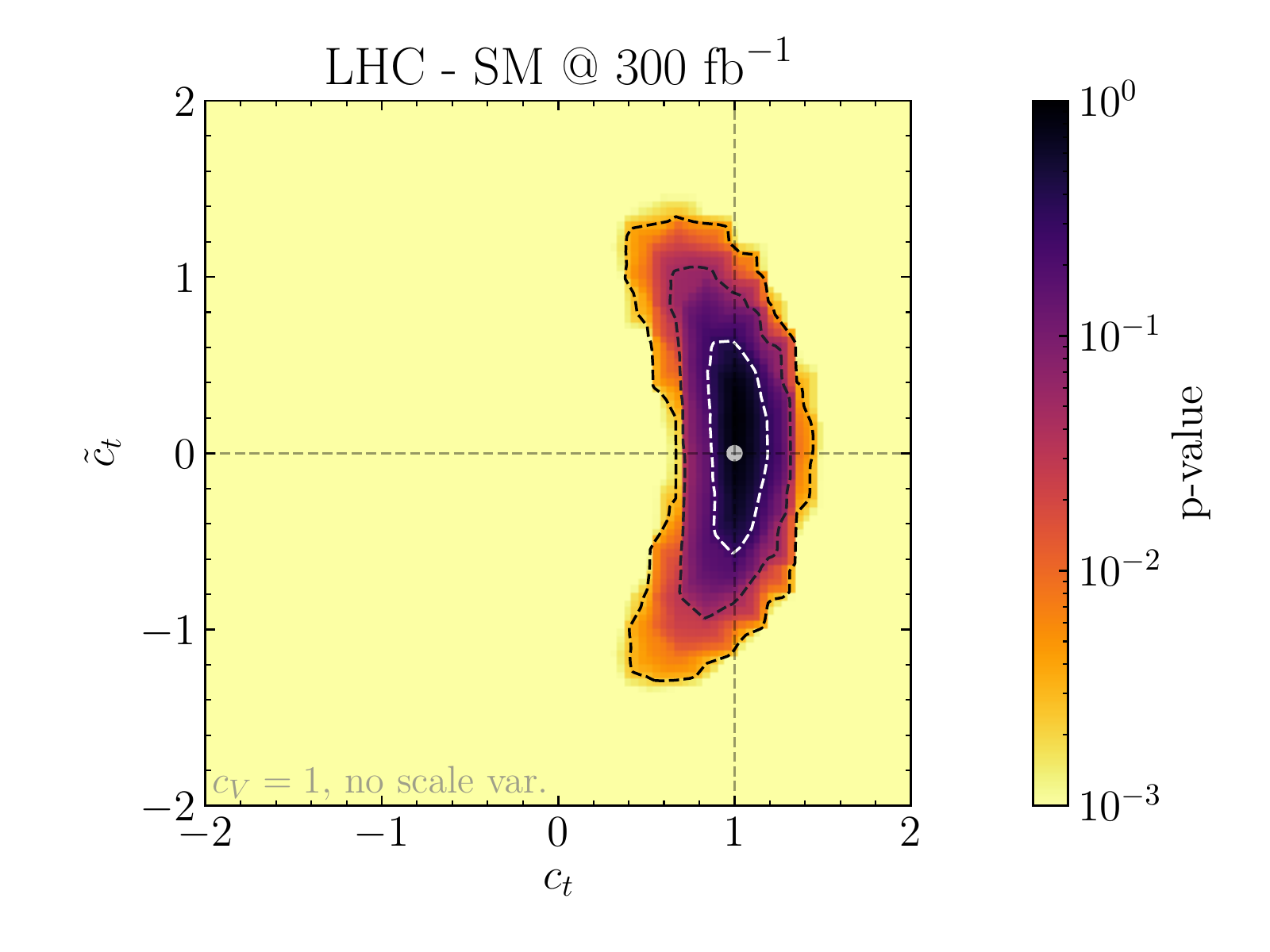}
    \includegraphics[width=0.49\textwidth]{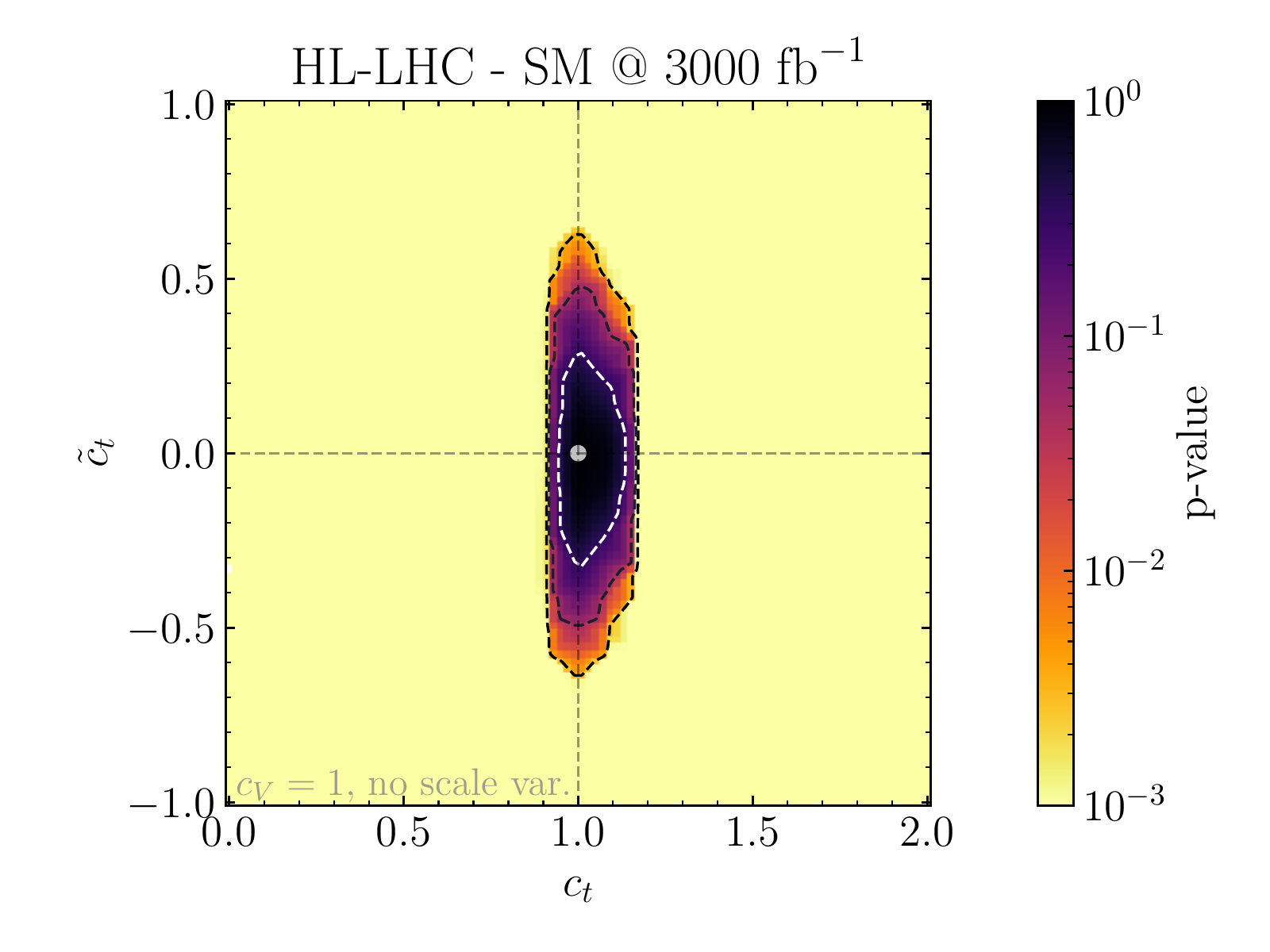}
    \includegraphics[width=0.49\textwidth]{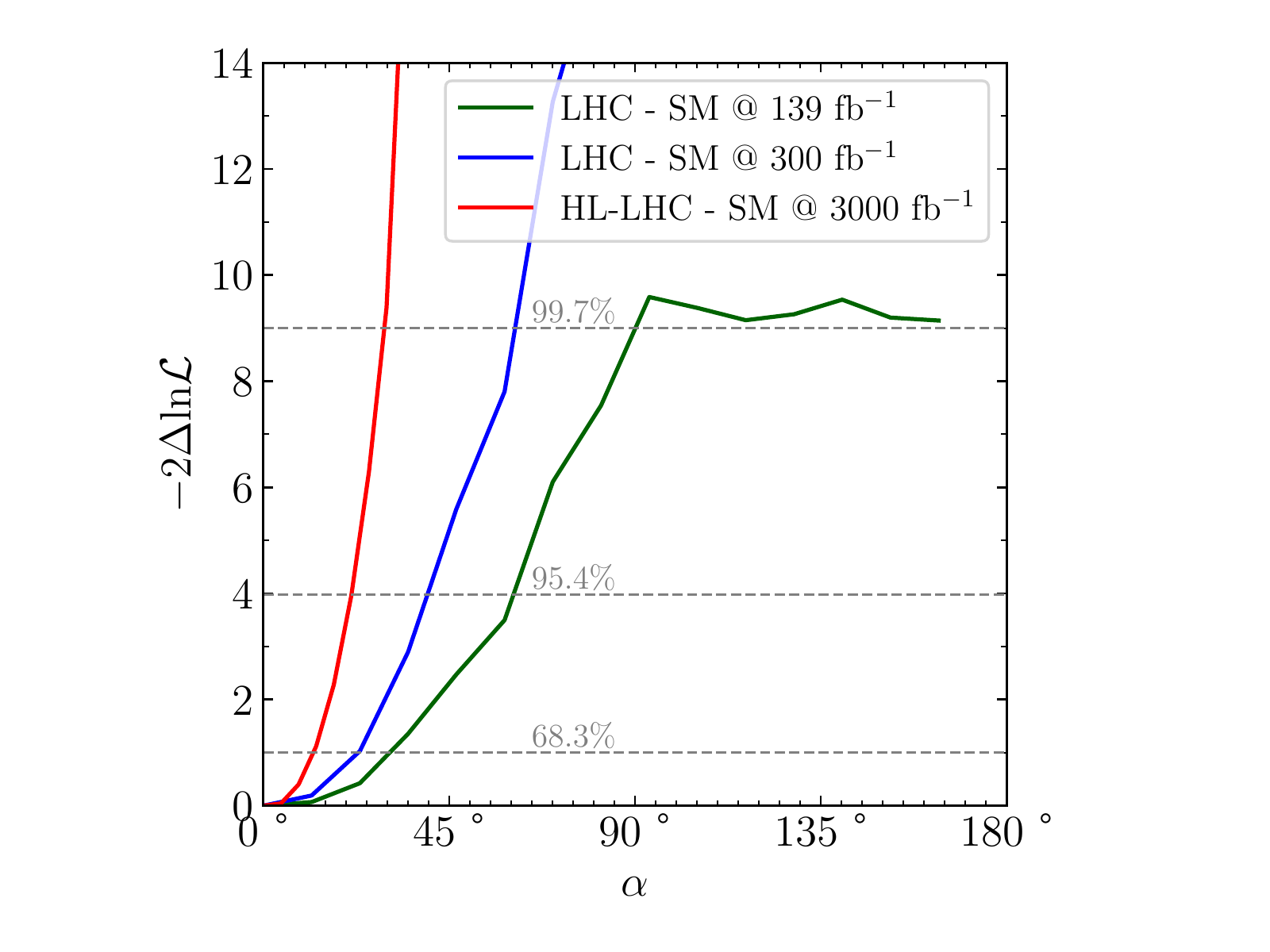}
    \caption{\textit{Upper left:} Expected limits (assuming SM data) on \ct and \cttilde using a luminosity of $139\invfb$. We assume that $\cv = 1$ and do not vary the renormalization scale. \textit{Upper right:} same as upper left panel but for a luminosity of $300\invfb$. \textit{Bottom left:} Same as upper left panel, but the constraints are evaluated for the HL-LHC with a luminosity of $3000\invfb$. \textit{Bottom right:} Constraints on the \cp-violating phase $\alpha$ for a luminosity of $139\invfb$ (green), $300\invfb$ (blue), and $3000\invfb$ (red) after profiling of $|g_t|$.}
    \label{fig:expected_ct_cttilde}
\end{figure}

The resulting limits in the $(\ct,\cttilde)$ plane are shown in the upper left, the upper right, and  the lower left panels of \cref{fig:expected_ct_cttilde}. The colour coding indicates the $p$-value associated with each parameter point. The white and black dashed contours define the $68.3\%$, $95.4\%$, and $99.7\%$ confidence level (CL) limits, respectively. For these plots, we project the three-dimensional likelihood to the $(\ct,\cttilde)$ plane by setting $\cv = 1$ and we do not perform any variation of the renormalization scale here.

In the upper left plot of \cref{fig:expected_ct_cttilde}, the constraints are shown for a luminosity of $139\invfb$. With this amount of data, the negative \ct range cannot be completely excluded at the $99.7\%$ CL. The form of the $99.7\%$ CL region is a consequence of the dependence of the top-associated Higgs production cross section on \ct and \cttilde (see e.g.\ \ccite{Bahl:2020wee}). With $139\invfb$, \cttilde is constrained to the interval $\sim [-0.8,0.8]$ at the $68.3\%$ CL level. Increasing the luminosity to $300\invfb$ (see upper right plot of \cref{fig:expected_ct_cttilde}) tightens the bounds on \cttilde to $\sim [-0.5,0.5]$ and excludes the negative \ct region at the $99.7\%$ CL level. Even stronger constraints are possible at the HL-LHC (see bottom left plot of \cref{fig:expected_ct_cttilde}; note the reduced parameter range) tentatively constraining \cttilde to the interval $\sim [-0.25,0.25]$. At the HL-LHC, the analysis does not only profit from the increased luminosity but also from the improved detector coverage in the forward region. While the bound on \cttilde for $\mathcal{L}=300\invfb$ corresponds roughly to a rescaling of the bound for $\mathcal{L}=139\invfb$ by the increased luminosity, the improvement at the HL-LHC is weaker than expected by a simple luminosity rescaling. This is a consequence of the small dependence of the total rate on \cttilde close to the SM point. In this situation, the constraints on \cttilde rely increasingly on the information encoded in the kinematic distributions.

These results can also be presented in terms of a \cp-violating phase $\alpha$ and an absolute value $|g_t|$ of the top-Yukawa coupling (see \cref{eq:alpha_gt}). Profiling over $|g_t|$, we show the one-dimensional profiles for $\alpha$ in the lower right panel of \cref{fig:expected_ct_cttilde} for the LHC using $139\invfb$ (green line), the LHC using $300\invfb$ (blue line), and the HL-LHC using $3000\invfb$ (red line).\footnote{Here, we directly show the negative likelihood. In the other Figures, we display the $p$-value obtained using Wilks' theorem.} With a luminosity of $139\invfb$, $\alpha$ is hardly constrained $\alpha$ at the 99.7\% confidence level. At 95.4\% confidence level, $\alpha$ is constrained to be below $\sim 60^\circ$, $\sim 40^\circ$, and $\sim 22^\circ$ using $139\invfb$, $300\invfb$, and $3000 \invfb$ respectively.

The expected bound on $\alpha$ using $139\invfb$ are close to the precision level reached in \ccite{Sirunyan:2020sum,Aad:2020ivc,CMS:2020dkv,CMS:2021nnc}, which also focused on top-associated Higgs production. The experimental analyses, however, also consider events which do not contain a lepton.\footnote{Note moreover that the effective model used in \ccite{Aad:2020ivc} is not directly comparable to the model used in this work (see \cref{sec:model}), since MC events have been generated at next-to-leading order (NLO) implying an interdependence of Higgs production via gluon fusion and top-associated Higgs production. Moreover, the results of \ccite{CMS:2020dkv,CMS:2021nnc} are not directly comparable, since two Higgs decay channels ($H\to \gamma\gamma$ and $H\to 4\ell$) have been combined.} Note also that for a luminosity of $139\invfb$, the constraints on \cttilde are mainly due to measurements of the total rate (as indicated by the ellipse-shaped $99.7\%$ CL allowed region in the upper left plot of \cref{fig:expected_ct_cttilde}). With an increased luminosity, kinematic constraints become increasingly important. Correspondingly, we expect machine-learning-based inference to demonstrate its advantages more clearly if the luminosity is increased.

\medskip

\begin{figure}
    \centering
    \includegraphics[width=0.48\textwidth]{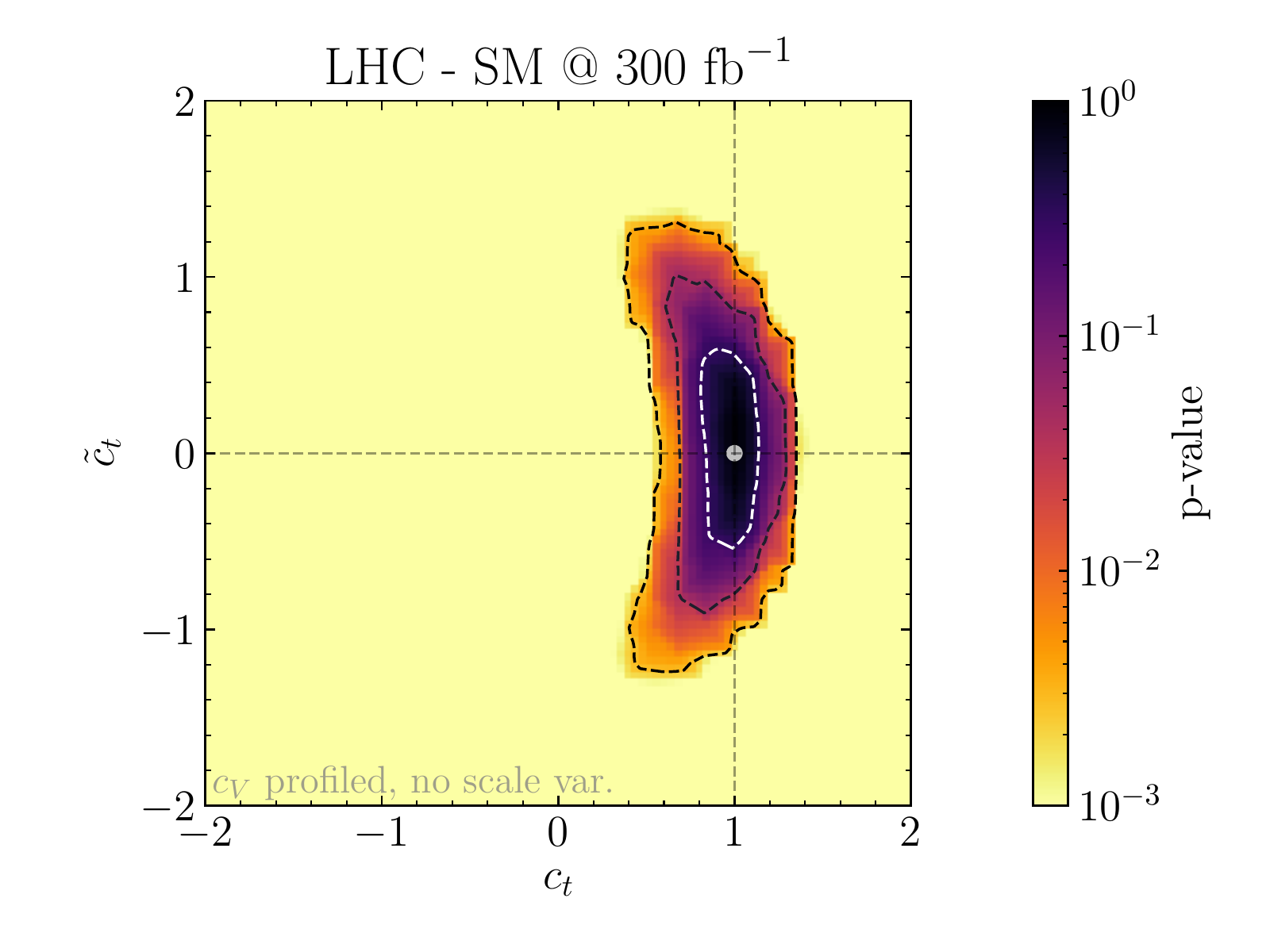}
    \includegraphics[width=0.48\textwidth]{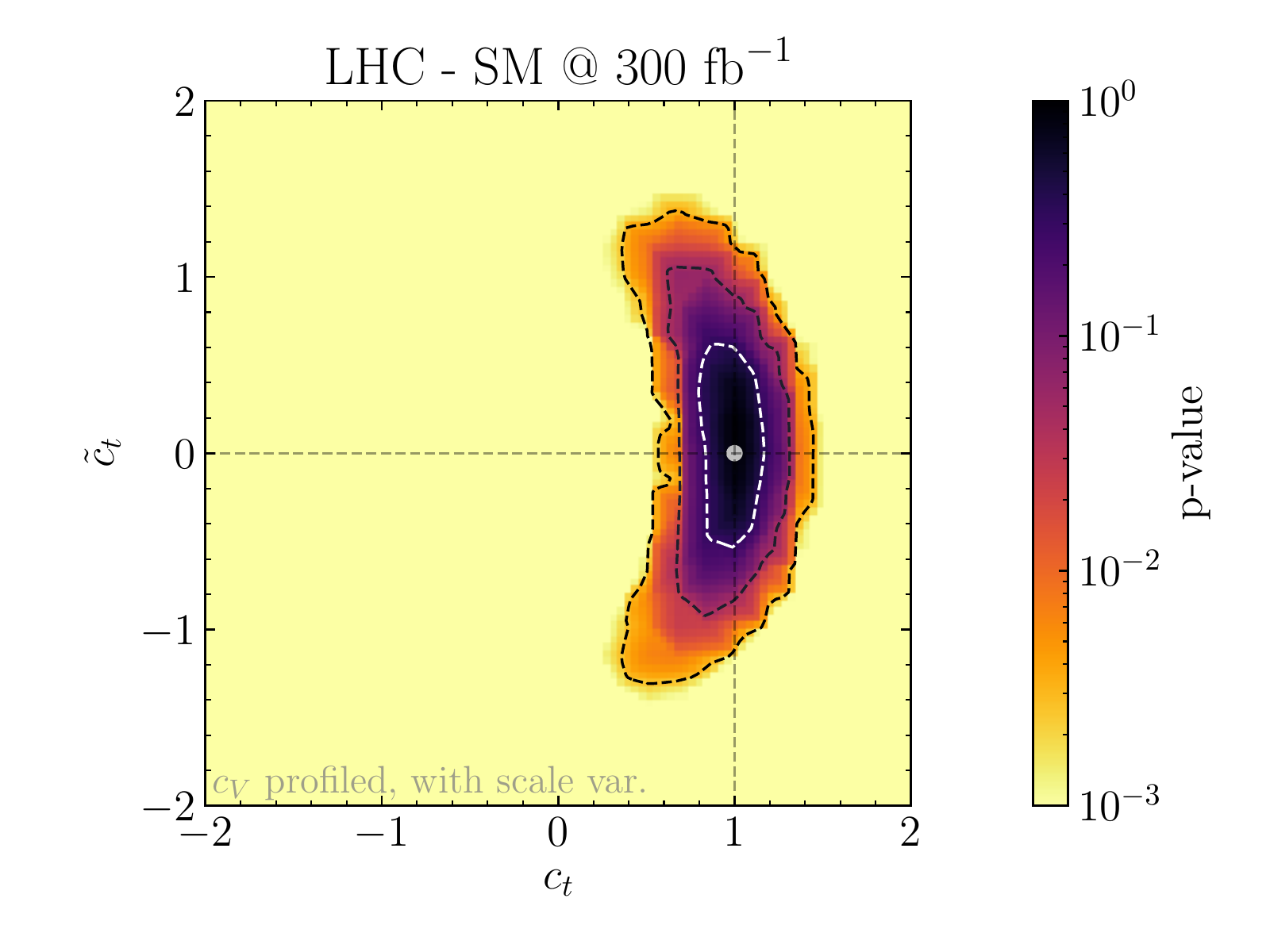}
    \caption{\textit{Left:} Expected limits on \ct and \cttilde with a luminosity of $300\invfb$ for a freely floating \cv. \textit{Right:} Same as left, but also the renormalization scale is floated.}
    \label{fig:expected_ct_cttilde_profile}
\end{figure}

So far, we have assumed that the Higgs coupling to massive vector bosons is SM-like (i.e., $\cv = 1$). In order to assess the impact of \cv on the $(\ct, \cttilde)$ constraints, we now include \cv fully in our likelihood estimation (floating it freely in the interval $[0.5,1.5]$). The constraints on \ct and \cttilde (using a luminosity of $300\invfb$) after profiling over \cv are shown in the left panel of \cref{fig:expected_ct_cttilde_profile}. In comparison to the upper right panel of \cref{fig:expected_ct_cttilde}, for which $\cv = 1$ is fixed, the constraints in the $(\ct,\cttilde)$ plane are only slightly weaker (i.e., the 95.4\% C.L. limit on $\alpha$ is weakened to $\sim 45^\circ$). This indicates that even without a precise knowledge of the Higgs coupling to massive vector boson, top-associated Higgs production allows for a precise determination of the top-Yukawa coupling.

An additional source of uncertainty not considered so far are uncertainties of theoretical nature. In order to estimate them, we consider a variation of the renormalization (and factorization) scale entering the MC event generation. The resulting constraints on \ct and \cttilde (using a luminosity of $300\invfb$) after profiling over the scale nuisance parameter (and $\cv$) are shown in the right panel of \cref{fig:expected_ct_cttilde_profile}. Also here, the constraints are only slightly weaker than those presented in the upper right panel of \cref{fig:expected_ct_cttilde} and the left panel of \cref{fig:expected_ct_cttilde_profile} (finding a 95.4\% CL limit of $\sim 45^\circ$ on $\alpha$). This result indicates that our result is relatively robust against theoretical uncertainties even though a more detailed study taking into account e.g.\ PDF uncertainties as well as next-to-leading order effects, which is beyond the scope of the present paper, would be necessary to answer this question more quantitatively.


\subsection{Observed limits in case of a deviation from the SM}
\label{sec:res_CPmixed}

All results presented above show expected limits assuming SM data. It is, however, also relevant to ask how well \ct and \cttilde can be constrained if the data is not SM like.

\begin{figure}
    \centering
    \includegraphics[width=0.48\textwidth]{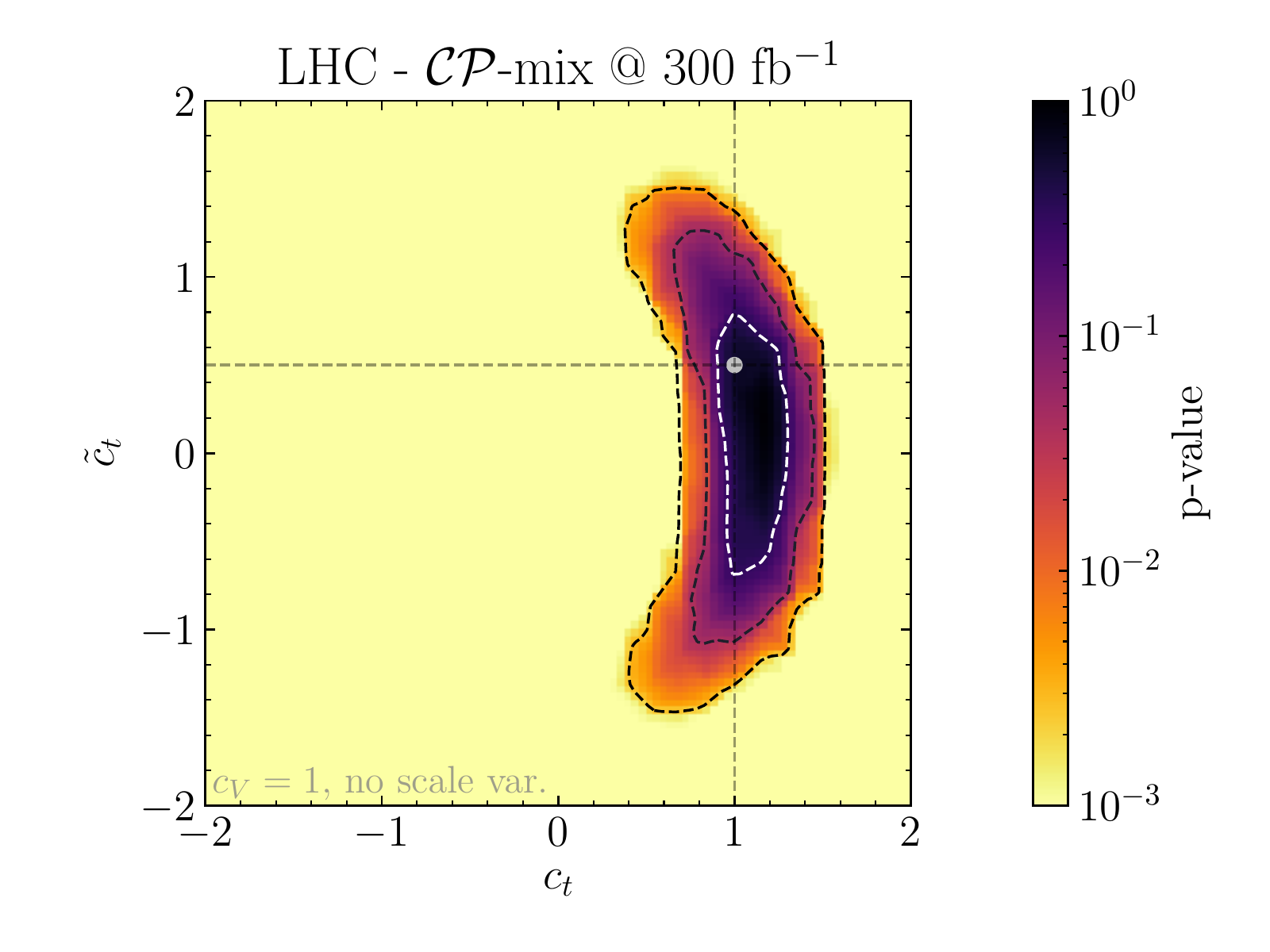}
    \includegraphics[width=0.48\textwidth]{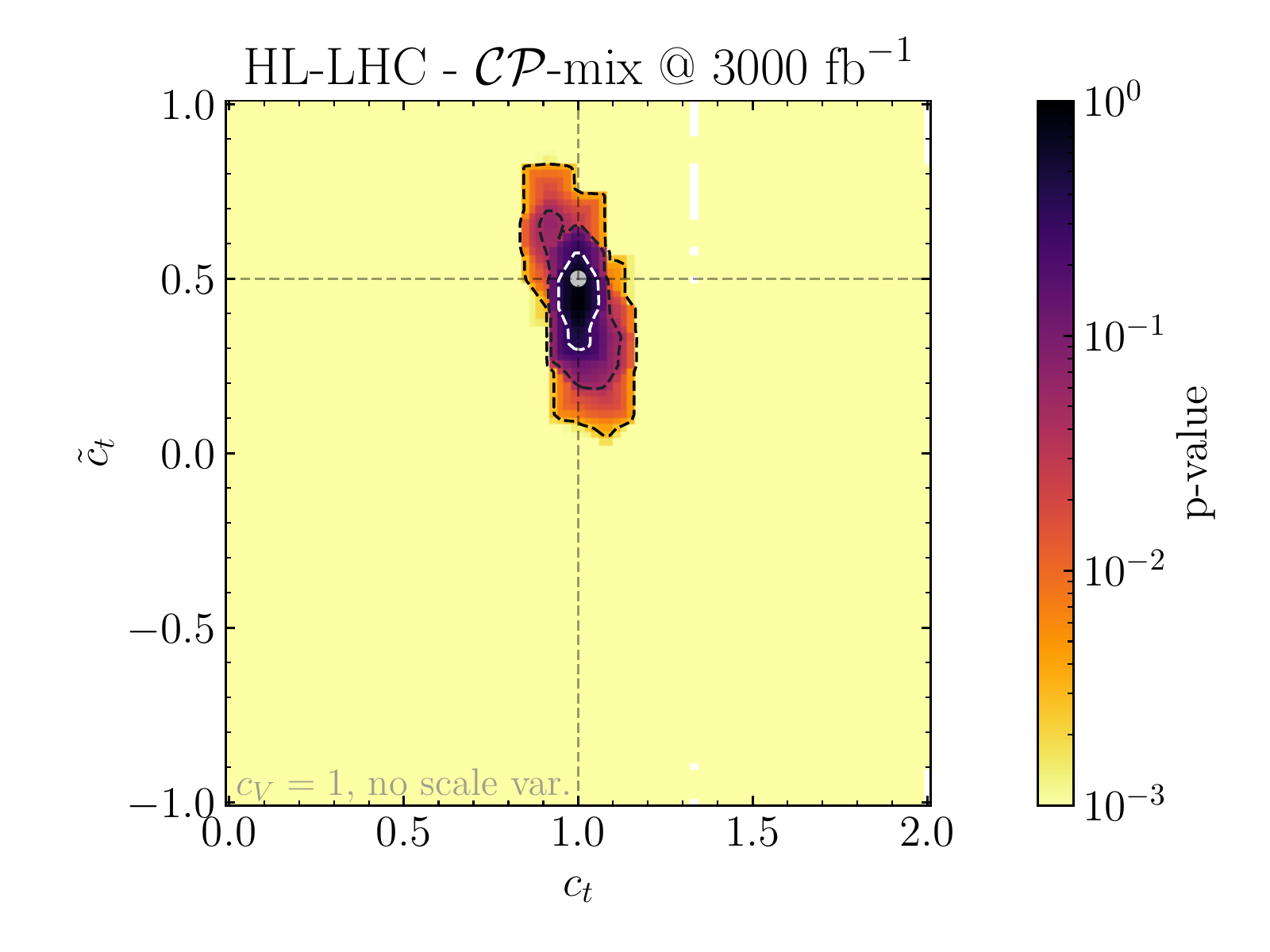}
    \caption{Observed limit on \ct and \cttilde for a \cp-mixed Higgs boson (with $\ct = 1$ and $\cttilde = 0.5$) using a luminosity of $300\invfb$ (left) and $3000\invfb$ (right).}
    \label{fig:observed_ct_cttilde}
\end{figure}

We address this question in \cref{fig:observed_ct_cttilde} where we assume the observed data to follow the predictions for a \cp-admixed Higgs boson. For this \cp-admixed Higgs boson, we fix $\ct = 1$ and $\cttilde = 0.5$ as an example. Moreover, we assume that $\cv = 1$.

In the left panel of \cref{fig:observed_ct_cttilde}, showing the exclusion boundaries for a luminosity of $300\invfb$ collected at the LHC, we observe that the SM hypothesis cannot be excluded even at the $68.3\%$ CL level. In comparison to the expected exclusion boundaries in case of SM data (see \cref{fig:expected_ct_cttilde}), the $68.3\%$ CL allowed \cttilde interval weakens by $\sim 50\%$.

The situation is completely different for a luminosity of $3000\invfb$ collected at HL-LHC (see right panel of \cref{fig:observed_ct_cttilde}). Here, the SM point is excluded at the $99.7\%$ CL level. The HL-LHC precision level allows to quite precisely pinpoint \ct (to $\sim [0.9,1.1]$) and \cttilde (to $\sim [0.3,0,6]$) at the $68.3\%$ CL level.


\subsection{Most sensitive observables}
\label{sec:resuls_obs}

After discussing the expected constraints on \ct and \cttilde, it is a relevant question which observables are most important for setting these limits. We approach this question by evaluating the Fisher information for different sets of input observables for a luminosity of $300\invfb$. 

As a first step, we evaluate the Fisher information matrix for the SM parameter point ($\cv = \ct = 1$, $\cttilde = 0$) taking into account the full kinematic and cross-section information (labelled by ``full''),\footnote{For evaluating the kinematic information, we rely on an ensemble of three \texttt{SALLY} estimators trained with a sample of $10^5$ unweighted events at the SM and the \cp-mixed benchmark points.}
\begin{align}\label{eq:Iij_SM}
    I_{ij}^\text{full}(\text{SM}) \simeq
    \begin{pmatrix}
        91.4 &  13.7 &  0.1 \\
        13.7 & 108.2 & -0.1 \\
         0.1 &  -0.1 &  0.004
    \end{pmatrix},
\end{align}
where we span our parameter space by the vector $(\cv, \ct, \cttilde)^T$. As eigenvalues of $I_{ij}^\text{full}(\text{SM})$, we obtain $115.9$, $83.7$, and $0.001$ with the respective normalized eigenvectors
\begin{align}\label{eq:Iij_evecs_SM}
    v_1 \simeq
    \begin{pmatrix}
        0.49 \\ 0.87 \\ -0.001
    \end{pmatrix}, \hspace{.5cm}
    v_2 \simeq
    \begin{pmatrix}
        0.87 \\ -0.49 \\ 0.001
    \end{pmatrix}, \hspace{.5cm}
    v_3 \simeq
    \begin{pmatrix}
        -0.0001 \\ 0.0013 \\ 1.00
    \end{pmatrix}.     
\end{align}
As explained in \cref{sec:fisher}, the Fisher information matrix is directly related to the reachable precision level for the various theory parameters: the higher the information, the more precise the corresponding coupling can be measured. Correspondingly, \cref{eq:Iij_SM,eq:Iij_evecs_SM} imply that close to the SM point \ct can be measured most precisely. There, however, is a strong correlation with \cv (as indicated by the comparably large off-diagonal entries), which can be constrained a little less. In contrast, the sensitivity to \cttilde is much lower. This finding is in agreement with the shape of the exclusion boundaries found in \cref{fig:expected_ct_cttilde} which are almost flat in the \cttilde direction close to the SM point.

\begin{figure}
    \centering
    \includegraphics[width=1\textwidth]{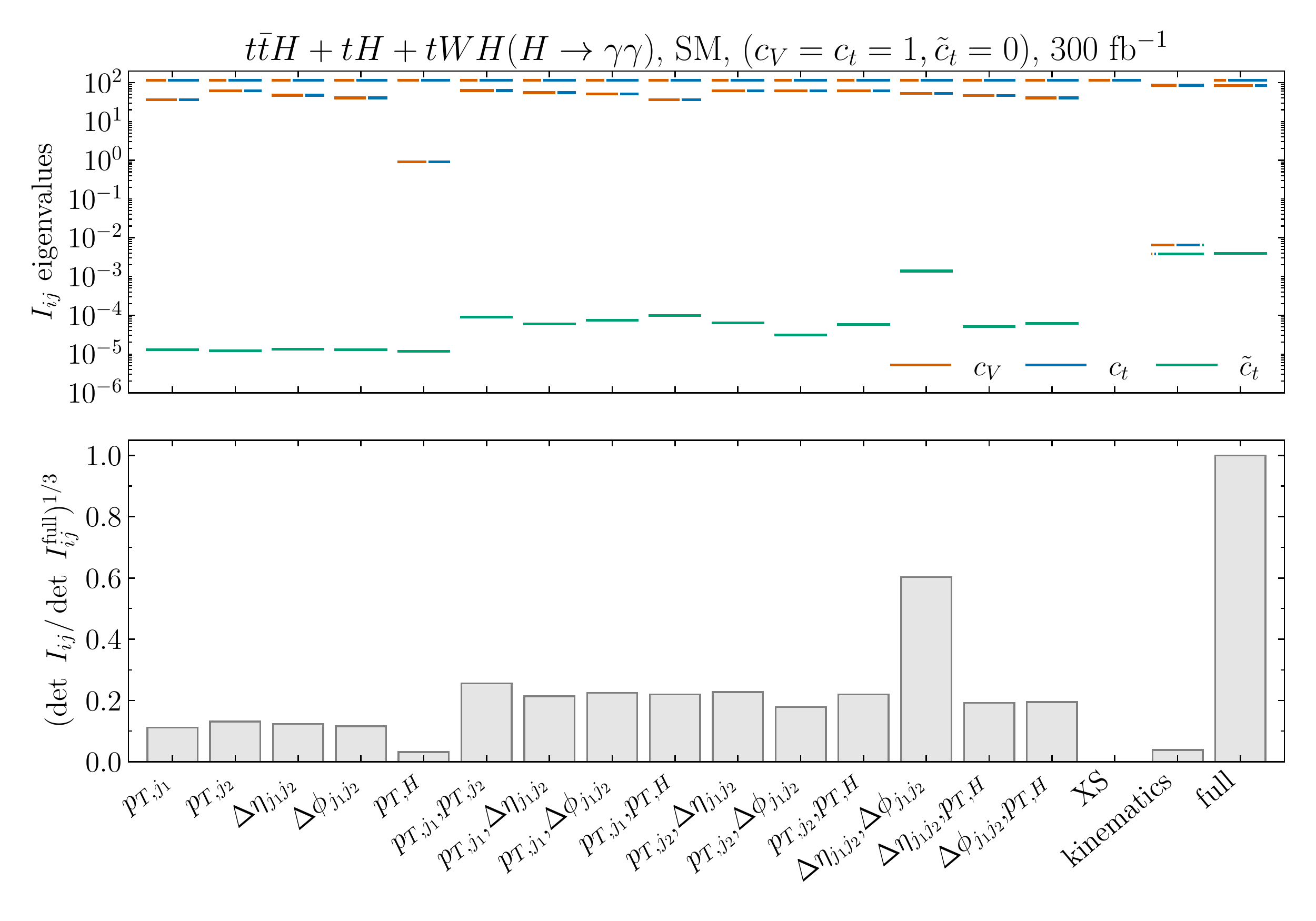}
    \caption{Fisher information for top-associated Higgs production with $H\to\gamma\gamma$ (for a luminosity of $300\invfb$). The full information is compared to the information contained in several one- and two-dimensional distributions. The Fisher information is evaluated for the SM benchmark point ($\cv = \ct = 1$, $\cttilde = 0$).}
    \label{fig:info_bar_SM}
\end{figure}

\begin{figure}
    \centering
    \includegraphics[width=1\textwidth]{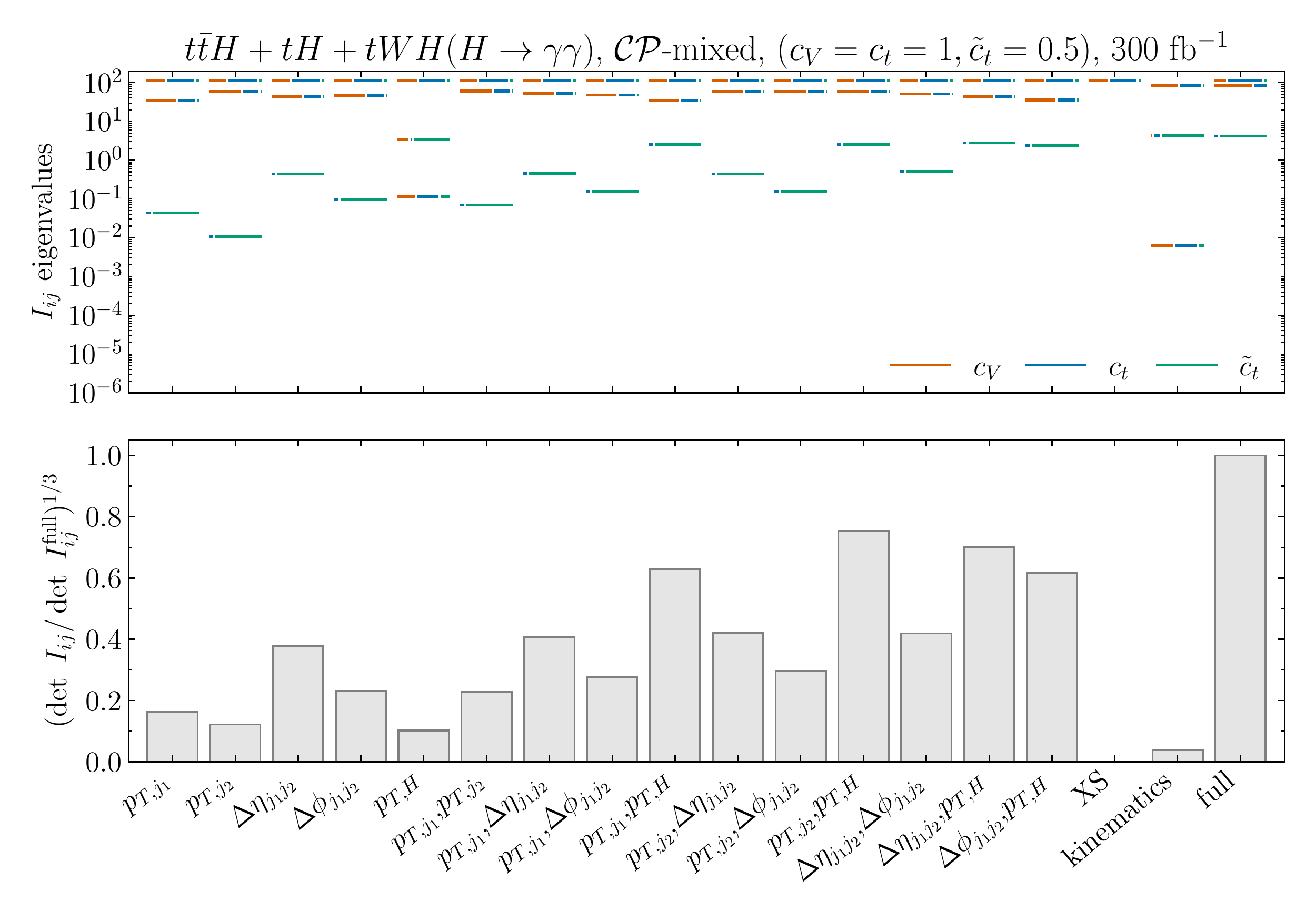}
    \caption{Same as \cref{fig:info_bar_SM} but the Fisher information is evaluated for a \cp-mixed benchmark point ($\cv = \ct = 1$, $\cttilde = 0.5$).}
    \label{fig:info_bar_CPmixed}
\end{figure}

As a second step, we recompute the Fisher information based on a selection of one- and two-dimensional kinematic distributions in order to understand which observables contain the most information about the various parameters.\footnote{We also evaluated the information contained in all other evaluated observables (see \cref{sec:event_simulation}) and their pair-wise combinations. In \cref{fig:info_bar_SM,fig:info_bar_CPmixed}, we only display those observables with the highest amount of information or those of special experimental interest. The one-dimensional distributions are evaluated by filling a histogram with 20 bins; the two-dimensional distributions, by filling a two-dimensional histogram with $10\times 10$ bins.} Moreover, we evaluate the Fisher information using only the kinematic information (labelled by ``kinematics'') and using only the cross-section information (labelled by ``XS''). Note that the one- and two-dimensional kinematic distributions always implicitly contain the cross section information.

In \cref{fig:info_bar_SM}, we show the results of the different evaluations of the Fisher information at the SM point. The panel is divided in two sub-panels. In the upper sub-panels, the eigenvalues of the Fisher matrix are shown. As explained for \cref{eq:Iij_SM,eq:Iij_evecs_SM}, the size and decomposition --- as indicated by the colour coding (orange for \cv, blue for \ct, green for \cttilde) --- of these eigenvalues is a measure which combinations of parameters can be constrained most strongly to a certain precision. In the lower sub-panels, the determinant of the Fisher matrix, which can be used as a measure of the overall reachable precision level, is shown. We normalize it to the determinant of the Fisher matrix calculated taking into account the full kinematic and cross-section information and take the third root to obtain a measure for the average precision level reachable for each of three parameters.

As already discussed above, the highest precision level is reachable for \ct. The precision level reachable for \cv is slightly smaller. As visible by comparing the different ways of evaluating the Fisher information, the information about \ct and \cv is to a large extent contained in the total rate. Taking into account only kinematic information results in significantly weaker constraints. Note, however, that the cross section alone is only sensitive to a specific combination of \ct and \cv, since the cross section is only a single number. Consequently, the determinant of the Fisher matrix calculated based only on the cross section is zero. If, however, the kinematic distribution of at least one observable is taken into account in addition to the total rate, the constraints on \ct and \cv are almost as strong as if the full information is considered. The situation is, however, quite different for constraints on \cttilde, which are in general much weaker than the constraints on \ct and \cv. The information on \cttilde is almost exclusively contained in the kinematic distributions as evident when comparing the eigenvalues of the Fisher matrix calculated based on the full and on only the kinematic information. This kinematic information on \cttilde can also hardly be captured by one- or two-dimensional kinematic distributions. Only the two-dimensional histogram of the pseudo-rapidity and angular difference of the leading and the subleading jet is able to capture a significant part of the information on \cttilde. In summary, it seems to be necessary to fully exploit all available information to constrain \cttilde at the SM point.

The situation is different if the Fisher information is evaluated at a \cp-mixed benchmark point ($\cv = \ct = 1$, $\cttilde = 0.5$, as for the \cp-admixed Higgs boson in \cref{sec:res_CPmixed}). Using the full kinematic and cross section information, we obtain for the Fisher information matrix
\begin{align}
    I_{ij}^\text{full}(\text{\cp-mixed}) \simeq
    \begin{pmatrix}
        91.3 &  10.6 &  6.4 \\
        10.6 &  98.2 & 26.3 \\
         6.4 &  26.3 & 11.7
    \end{pmatrix}
\end{align}
with the eigenvalues $112.4$, $84.6$, and $4.2$ as well as the corresponding eigenvectors
\begin{align}
    v_1 \simeq
    \begin{pmatrix}
        0.50 \\ 0.87 \\ -0.04
    \end{pmatrix}, \hspace{.5cm}
    v_2 \simeq
    \begin{pmatrix}
        0.83 \\ -0.49 \\ -0.27
    \end{pmatrix}, \hspace{.5cm} 
    v_3 \simeq
    \begin{pmatrix}
        0.25 \\ -0.10 \\ 0.96
    \end{pmatrix}.
\end{align}
While \ct can still be constrained most precisely (and still is strongly correlated with \cv), \cttilde can be constrained more precisely in comparison to the SM point (see \cref{eq:Iij_SM}) and is now weakly correlated with both \ct (and too lesser extent with \cv).

The information obtained at the \cp-mixed benchmark point using the various kinematic observables (and their combination) is visualized in \cref{fig:info_bar_CPmixed}. As visible in the upper sub-panel, the information on \cttilde is to a large extent contained in the kinematic information as for the SM point. In contrast to the SM point, however, the one- and two-dimensional kinematic distributions are able to capture this information. For the \cp-mixed benchmark point, the one-dimensional transverse momentum distribution of the Higgs boson is able to capture the information on \cttilde almost completely. This is especially interesting in the light of recent simplified-template cross section (STXS) measurements of the Higgs transverse momentum for top-associated Higgs production~\cite{ATLAS:2020pvn}.\footnote{In \ccite{ATLAS:2020pvn}, $t\bar tH$ and $tH$ events are, however, disentangled based upon the assumption of SM-like kinematics. As indicated e.g.\ in \ccite{Bahl:2020wee}, this assumption does not hold in the case of a non-vanishing \cp-odd component of the top-Yukawa coupling.}

%% file: sec_conclusions.tex
A detailed investigation of the top-Yukawa coupling is crucial to probe the \cp nature of the Higgs boson discovered at the LHC. In this work, we assessed the potential of machine-learning-based inference to constrain a possible \cp-odd component of the Higgs--top-quark interaction.

Machine-learning-based inference allows to approximate the full likelihood fully exploiting the kinematic information contained in the event data. It, moreover, fully takes into account parton shower and detector effects without relying on simplifying assumptions like e.g.\ the matrix element method.

We performed our analysis in a simplified model framework in which not only the top-Yukawa coupling is allowed to vary freely but also a global rescaling factor is introduced for the Higgs interactions with massive vector bosons.

As physical target process, we focused on top-associated Higgs production with the Higgs decaying to two photons. In order to suppress background, we demanded at least one lepton in the final state. The method can, however, also be used for more inclusive final states.

Applying machine-learning-based inference to top-associated Higgs production --- using the tool \texttt{Madminer} ---, we derived expected bounds on a \cp-violating top-Yukawa coupling for the LHC and HL-LHC. Assuming SM data, we found that a \cp-odd top Yukawa coupling can be constrained at the $68.3\%$ CL level to lie within $\sim [-0.8,0.8]$ using a luminosity of $139\invfb$ at the LHC; with a luminosity of $300\invfb$, this bound is tightened to $\sim [-0.5,0.5]$; and, with a luminosity of $3000\invfb$ at the HL-LHC, a bound of $\sim [-0.25,0.25]$ can be obtained. The constraints for a luminosity of $139\invfb$ are of a similar precision as current experimental studies, which use a larger data set (i.e.~they also include events without a lepton in the final state). Assuming the existence of a small \cp-odd Yukawa coupling ($\cttilde = 0.5$), we found that such a deviation from the SM can not be distinguished from the SM at LHC. To establish a deviation at $99.7\%$ CL level, the full data set to be accumulated at the HL-LHC is needed.

In addition to deriving expected bounds, we also studied the impact of a non-SM-like Higgs--vector-boson coupling on the top-Yukawa coupling constraints. We found deviations of the Higgs--vector-boson couplings from the SM to have negligible impact on the top-Yukawa coupling constraints showing the promising potential of top-associated Higgs production for precision constraints on the top-Yukawa coupling. Moreover, we studied the impact of theoretical uncertainties by treating the renormalization scale as a nuisance parameter. We found also the variation of the renormalization scale to have small effect on our results indicating that our result is robust against theoretical uncertainties.

Aside of direct constraints on the \cp nature of the top-Yukawa coupling, we investigated which kinematic observables are most sensitive to a \cp-odd top-Yukawa coupling by extracting the Fisher information. While it is in general very hard to constrain a \cp-odd top-Yukawa coupling if the SM is realized in nature, we found the transverse momentum distribution of the Higgs boson to have a high sensitivity to a \cp-odd top-Yukawa coupling if this coupling is non-zero.

Our analysis shows the potential of machine-learning based inference to probe the \cp character of the Higgs--top-quark interaction. We hope that the present study triggers future work aiming a exploiting the full kinematic information of top-quark-associated Higgs production.

%% file: bibliography.bib
@article{Freitas:2012kw,
      author         = "Freitas, Ayres and Schwaller, Pedro",
      title          = "{Higgs CP Properties From Early LHC Data}",
      journal        = "Phys. Rev.",
      volume         = "D87",
      year           = "2013",
      number         = "5",
      pages          = "055014",
      doi            = "10.1103/PhysRevD.87.055014",
      eprint         = "1211.1980",
      archivePrefix  = "arXiv",
      primaryClass   = "hep-ph",
      reportNumber   = "ANL-HEP-PR-12-84",
      SLACcitation   = "%%CITATION = ARXIV:1211.1980;%%"
}

@article{Artoisenet:2013puc,
      author         = "Artoisenet, P. and others",
      title          = "{A framework for Higgs characterisation}",
      journal        = "JHEP",
      volume         = "11",
      year           = "2013",
      pages          = "043",
      doi            = "10.1007/JHEP11(2013)043",
      eprint         = "1306.6464",
      archivePrefix  = "arXiv",
      primaryClass   = "hep-ph",
      reportNumber   = "CERN-PH-TH-2013-148",
      SLACcitation   = "%%CITATION = ARXIV:1306.6464;%%"
}

@article{Maltoni:2013sma,
      author         = "Maltoni, Fabio and Mawatari, Kentarou and Zaro, Marco",
      title          = "{Higgs characterisation via vector-boson fusion and
                        associated production: NLO and parton-shower effects}",
      journal        = "Eur. Phys. J.",
      volume         = "C74",
      year           = "2014",
      number         = "1",
      pages          = "2710",
      doi            = "10.1140/epjc/s10052-013-2710-5",
      eprint         = "1311.1829",
      archivePrefix  = "arXiv",
      primaryClass   = "hep-ph",
      SLACcitation   = "%%CITATION = ARXIV:1311.1829;%%"
}

@article{Demartin:2014fia,
      author         = "Demartin, Federico and Maltoni, Fabio and Mawatari,
                        Kentarou and Page, Ben and Zaro, Marco",
      title          = "{Higgs characterisation at NLO in QCD: CP properties of
                        the top-quark Yukawa interaction}",
      journal        = "Eur. Phys. J.",
      volume         = "C74",
      year           = "2014",
      number         = "9",
      pages          = "3065",
      doi            = "10.1140/epjc/s10052-014-3065-2",
      eprint         = "1407.5089",
      archivePrefix  = "arXiv",
      primaryClass   = "hep-ph",
      reportNumber   = "CP3-14-59, LPN14-096, MCNET-14-21",
      SLACcitation   = "%%CITATION = ARXIV:1407.5089;%%"
}

@article{Demartin:2015uha,
      author         = "Demartin, Federico and Maltoni, Fabio and Mawatari,
                        Kentarou and Zaro, Marco",
      title          = "{Higgs production in association with a single top
                        quark at the LHC}",
      journal        = "Eur. Phys. J.",
      volume         = "C75",
      year           = "2015",
      number         = "6",
      pages          = "267",
      doi            = "10.1140/epjc/s10052-015-3475-9",
      eprint         = "1504.00611",
      archivePrefix  = "arXiv",
      primaryClass   = "hep-ph",
      reportNumber   = "MCNET-15-07, CP3-15-08",
      SLACcitation   = "%%CITATION = ARXIV:1504.00611;%%"
}

@article{Demartin:2016axk,
      author         = "Demartin, Federico and Maier, Benedikt and Maltoni, Fabio
                        and Mawatari, Kentarou and Zaro, Marco",
      title          = "{tWH associated production at the LHC}",
      journal        = "Eur. Phys. J.",
      volume         = "C77",
      year           = "2017",
      number         = "1",
      pages          = "34",
      doi            = "10.1140/epjc/s10052-017-4601-7",
      eprint         = "1607.05862",
      archivePrefix  = "arXiv",
      primaryClass   = "hep-ph",
      reportNumber   = "MCNET-16-30, CP3-16-40, LPSC16158",
      SLACcitation   = "%%CITATION = ARXIV:1607.05862;%%"
}

@article{Azevedo:2017qiz,
      author         = "Azevedo, D. and Onofre, A. and Filthaut, F. and Gonçalo,
                        R.",
      title          = "{CP tests of Higgs couplings in $t\bar{t}h$ semileptonic
                        events at the LHC}",
      journal        = "Phys. Rev.",
      volume         = "D98",
      year           = "2018",
      number         = "3",
      pages          = "033004",
      doi            = "10.1103/PhysRevD.98.033004",
      eprint         = "1711.05292",
      archivePrefix  = "arXiv",
      primaryClass   = "hep-ph",
      SLACcitation   = "%%CITATION = ARXIV:1711.05292;%%"
}

@article{Barger:2018tqn,
      author         = "Barger, Vernon and Hagiwara, Kaoru and Zheng, Ya-Juan",
      title          = "{Probing the Higgs Yukawa coupling to the top quark at
                        the LHC via single top+Higgs production}",
      journal        = "Phys. Rev.",
      volume         = "D99",
      year           = "2019",
      number         = "3",
      pages          = "031701",
      doi            = "10.1103/PhysRevD.99.031701",
      eprint         = "1807.00281",
      archivePrefix  = "arXiv",
      primaryClass   = "hep-ph",
      reportNumber   = "KEK-TH-2058, PITT-PACC-1811, OU-HET-974",
      SLACcitation   = "%%CITATION = ARXIV:1807.00281;%%"
}

@article{Cao:2019ygh,
      author         = "Cao, Qing-Hong and Chen, Shao-Long and Liu, Yandong and
                        Zhang, Rui and Zhang, Ya",
      title          = "{Limiting top quark-Higgs boson interaction and
                        Higgs-boson width from multitop productions}",
      journal        = "Phys. Rev.",
      volume         = "D99",
      year           = "2019",
      number         = "11",
      pages          = "113003",
      doi            = "10.1103/PhysRevD.99.113003",
      eprint         = "1901.04567",
      archivePrefix  = "arXiv",
      primaryClass   = "hep-ph",
      SLACcitation   = "%%CITATION = ARXIV:1901.04567;%%"
}

@article{Goncalves:2018agy,
      author         = "Gonçalves, Dorival and Kong, Kyoungchul and Kim, Jeong
                        Han",
      title          = "{Probing the top-Higgs Yukawa CP structure in dileptonic
                        $ t\overline{t}h $ with M$_{2}$-assisted reconstruction}",
      journal        = "JHEP",
      volume         = "06",
      year           = "2018",
      pages          = "079",
      doi            = "10.1007/JHEP06(2018)079",
      eprint         = "1804.05874",
      archivePrefix  = "arXiv",
      primaryClass   = "hep-ph",
      reportNumber   = "PITT-PACC-1807",
      SLACcitation   = "%%CITATION = ARXIV:1804.05874;%%"
}

@article{He:2014xla,
      author         = "He, Xiao-Gang and Li, Guan-Nan and Zheng, Ya-Juan",
      title          = "{Probing Higgs boson $CP$ Properties with $t\bar{t}H$ at
                        the LHC and the 100 TeV $pp$ collider}",
      journal        = "Int. J. Mod. Phys.",
      volume         = "A30",
      year           = "2015",
      number         = "25",
      pages          = "1550156",
      doi            = "10.1142/S0217751X15501560",
      eprint         = "1501.00012",
      archivePrefix  = "arXiv",
      primaryClass   = "hep-ph",
      SLACcitation   = "%%CITATION = ARXIV:1501.00012;%%"
}

@article{Ellis:2013yxa,
      author         = "Ellis, John and Hwang, Dae Sung and Sakurai, Kazuki and
                        Takeuchi, Michihisa",
      title          = "{Disentangling Higgs-Top Couplings in Associated
                        Production}",
      journal        = "JHEP",
      volume         = "04",
      year           = "2014",
      pages          = "004",
      doi            = "10.1007/JHEP04(2014)004",
      eprint         = "1312.5736",
      archivePrefix  = "arXiv",
      primaryClass   = "hep-ph",
      reportNumber   = "KCL-PH-TH-2013-47, LCTS-2013-35, CERN-PH-TH-2013-312",
      SLACcitation   = "%%CITATION = ARXIV:1312.5736;%%"
}

@article{Brod:2013cka,
      author         = "Brod, Joachim and Haisch, Ulrich and Zupan, Jure",
      title          = "{Constraints on CP-violating Higgs couplings to the third
                        generation}",
      journal        = "JHEP",
      volume         = "11",
      year           = "2013",
      pages          = "180",
      doi            = "10.1007/JHEP11(2013)180",
      eprint         = "1310.1385",
      archivePrefix  = "arXiv",
      primaryClass   = "hep-ph",
      reportNumber   = "NSF-KITP-13-229",
      SLACcitation   = "%%CITATION = ARXIV:1310.1385;%%"
}

@article{Djouadi:2013qya,
      author         = "Djouadi, Abdelhak and Moreau, Grégory",
      title          = "{The couplings of the Higgs boson and its CP properties
                        from fits of the signal strengths and their ratios at the
                        7+8 TeV LHC}",
      journal        = "Eur. Phys. J.",
      volume         = "C73",
      year           = "2013",
      number         = "9",
      pages          = "2512",
      doi            = "10.1140/epjc/s10052-013-2512-9",
      eprint         = "1303.6591",
      archivePrefix  = "arXiv",
      primaryClass   = "hep-ph",
      reportNumber   = "LPT-ORSAY-13-19",
      SLACcitation   = "%%CITATION = ARXIV:1303.6591;%%"
}

@article{Alwall:2014hca,
      author         = "Alwall, J. and Frederix, R. and Frixione, S. and Hirschi,
                        V. and Maltoni, F. and Mattelaer, O. and Shao, H. -S. and
                        Stelzer, T. and Torrielli, P. and Zaro, M.",
      title          = "{The automated computation of tree-level and
                        next-to-leading order differential cross sections, and
                        their matching to parton shower simulations}",
      journal        = "JHEP",
      volume         = "07",
      year           = "2014",
      pages          = "079",
      doi            = "10.1007/JHEP07(2014)079",
      eprint         = "1405.0301",
      archivePrefix  = "arXiv",
      primaryClass   = "hep-ph",
      reportNumber   = "CERN-PH-TH-2014-064, CP3-14-18, LPN14-066, MCNET-14-09,
                        ZU-TH-14-14",
      SLACcitation   = "%%CITATION = ARXIV:1405.0301;%%"
}

@article{Sjostrand:2007gs,
      author         = "Sjostrand, Torbjorn and Mrenna, Stephen and Skands, Peter
                        Z.",
      title          = "{A Brief Introduction to PYTHIA 8.1}",
      journal        = "Comput. Phys. Commun.",
      volume         = "178",
      year           = "2008",
      pages          = "852-867",
      doi            = "10.1016/j.cpc.2008.01.036",
      eprint         = "0710.3820",
      archivePrefix  = "arXiv",
      primaryClass   = "hep-ph",
      reportNumber   = "CERN-LCGAPP-2007-04, LU-TP-07-28,
                        FERMILAB-PUB-07-512-CD-T",
      SLACcitation   = "%%CITATION = ARXIV:0710.3820;%%"
}

@article{Boudjema:2015nda,
      author         = "Boudjema, Fawzi and Godbole, Rohini M. and Guadagnoli,
                        Diego and Mohan, Kirtimaan A.",
      title          = "{Lab-frame observables for probing the top-Higgs
                        interaction}",
      journal        = "Phys. Rev.",
      volume         = "D92",
      year           = "2015",
      number         = "1",
      pages          = "015019",
      doi            = "10.1103/PhysRevD.92.015019",
      eprint         = "1501.03157",
      archivePrefix  = "arXiv",
      primaryClass   = "hep-ph",
      reportNumber   = "LAPTH-004-15, MSUHEP-150113",
      SLACcitation   = "%%CITATION = ARXIV:1501.03157;%%"
}

@article{deFlorian:2016spz,
      author         = "de Florian, D. and others",
      title          = "{Handbook of LHC Higgs Cross Sections: 4. Deciphering the
                        Nature of the Higgs Sector}",
      collaboration  = "LHC Higgs Cross Section Working Group",
      doi            = "10.2172/1345634, 10.23731/CYRM-2017-002",
      year           = "2016",
      eprint         = "1610.07922",
      archivePrefix  = "arXiv",
      primaryClass   = "hep-ph",
      reportNumber   = "FERMILAB-FN-1025-T, CERN-2017-002-M",
      SLACcitation   = "%%CITATION = ARXIV:1610.07922;%%"
}

@article{Kraus:2019myc,
      author         = "Kraus, Manfred and Martini, Till and Peitzsch, Sascha and
                        Uwer, Peter",
      title          = "{Exploring BSM Higgs couplings in single top-quark
                        production}",
      year           = "2019",
      eprint         = "1908.09100",
      archivePrefix  = "arXiv",
      primaryClass   = "hep-ph",
      reportNumber   = "HU-EP-19/21",
      SLACcitation   = "%%CITATION = ARXIV:1908.09100;%%"
}

@Booklet{ATL-PHYS-PUB-2014-021,
    author         = "{ATLAS Collaboration}",
    title          = "{ATLAS Pythia~8 tunes to \(7~\text{TeV}\) data}",
    howpublished   = "{ATL-PHYS-PUB-2014-021}",
    url            = "https://cds.cern.ch/record/1966419",
    year           = "2014",
}

@article{deFavereau:2013fsa,
      author         = "de Favereau, J. and Delaere, C. and Demin, P. and
                        Giammanco, A. and Lemaître, V. and Mertens, A. and
                        Selvaggi, M.",
      title          = "{DELPHES 3, A modular framework for fast simulation of a
                        generic collider experiment}",
      collaboration  = "DELPHES 3",
      journal        = "JHEP",
      volume         = "02",
      year           = "2014",
      pages          = "057",
      doi            = "10.1007/JHEP02(2014)057",
      eprint         = "1307.6346",
      archivePrefix  = "arXiv",
      primaryClass   = "hep-ex",
      SLACcitation   = "%%CITATION = ARXIV:1307.6346;%%"
}

@article{Martin:2009iq,
    author = "Martin, A.D. and Stirling, W.J. and Thorne, R.S. and Watt, G.",
    archivePrefix = "arXiv",
    doi = "10.1140/epjc/s10052-009-1072-5",
    eprint = "0901.0002",
    journal = "Eur.\ Phys.\ J.\ C",
    pages = "189--285",
    primaryClass = "hep-ph",
    reportNumber = "IPPP-08-95, DCPT-08-190, CAVENDISH-HEP-08-16",
    title = "{Parton distributions for the LHC}",
    volume = "63",
    year = "2009"
}

@inproceedings{Whalley:2005nh,
    author = "Whalley, M.R. and Bourilkov, D. and Group, R.C.",
    archivePrefix = "arXiv",
    booktitle = "{HERA and the LHC: A Workshop on the implications of HERA for LHC physics. Proceedings, Part B}",
    eprint = "hep-ph/0508110",
    month = "8",
    pages = "575--581",
    title = "{The Les Houches accord PDFs (LHAPDF) and LHAGLUE}",
    year = "2005"
}

@inbook{Cepeda:2019klc,
    author = "Cepeda, M. and others",
    archivePrefix = "arXiv",
    booktitle = "{Report on the Physics at the HL-LHC,and Perspectives for the HE-LHC}",
    doi = "10.23731/CYRM-2019-007.221",
    eprint = "1902.00134",
    month = "12",
    pages = "221--584",
    primaryClass = "hep-ph",
    reportNumber = "CERN-LPCC-2018-04",
    title = "{Report from Working Group 2}: {Higgs Physics at the HL-LHC and HE-LHC}",
    volume = "7",
    year = "2019"
}

@article{Huet:1994jb,
    author = "Huet, Patrick and Sather, Eric",
    archivePrefix = "arXiv",
    doi = "10.1103/PhysRevD.51.379",
    eprint = "hep-ph/9404302",
    journal = "Phys. Rev. D",
    pages = "379--394",
    reportNumber = "SLAC-PUB-6479",
    title = "{Electroweak baryogenesis and standard model CP violation}",
    volume = "51",
    year = "1995"
}

@article{Gavela:1993ts,
    author = "Gavela, M.B. and Hernandez, P. and Orloff, J. and Pene, O.",
    archivePrefix = "arXiv",
    doi = "10.1142/S0217732394000629",
    eprint = "hep-ph/9312215",
    journal = "Mod. Phys. Lett. A",
    pages = "795--810",
    reportNumber = "CERN-TH-7081-93, LPTHE-ORSAY-93-48, HUTP-93-A036, HD-THEP-93-45",
    title = "{Standard model CP violation and baryon asymmetry}",
    volume = "9",
    year = "1994"
}

@article{Sirunyan:2020sum,
    author = "Sirunyan, Albert M and others",
    collaboration = "CMS",
    title = "{Measurements of $\mathrm{t\bar{t}}H$ Production and the CP Structure of the Yukawa Interaction between the Higgs Boson and Top Quark in the Diphoton Decay Channel}",
    eprint = "2003.10866",
    archivePrefix = "arXiv",
    primaryClass = "hep-ex",
    reportNumber = "CMS-HIG-19-013, CERN-EP-2020-028",
    doi = "10.1103/PhysRevLett.125.061801",
    journal = "Phys. Rev. Lett.",
    volume = "125",
    number = "6",
    pages = "061801",
    year = "2020"
}

@article{Aad:2020ivc,
    author = "Aad, Georges and others",
    collaboration = "ATLAS",
    title = "{$CP$ Properties of Higgs Boson Interactions with Top Quarks in the $t\bar{t}H$ and $tH$ Processes Using $H \rightarrow \gamma\gamma$ with the ATLAS Detector}",
    eprint = "2004.04545",
    archivePrefix = "arXiv",
    primaryClass = "hep-ex",
    reportNumber = "CERN-EP-2020-046",
    doi = "10.1103/PhysRevLett.125.061802",
    journal = "Phys. Rev. Lett.",
    volume = "125",
    number = "6",
    pages = "061802",
    year = "2020"
}

@article{Aad:2012tfa,
      author         = "Aad, G. and others",
      title          = "{Observation of a new particle in the search for the
                        Standard Model Higgs boson with the ATLAS detector at the
                        LHC}",
      collaboration  = "ATLAS",
      journal        = "Phys. Lett.",
      volume         = "B716",
      year           = "2012",
      pages          = "1-29",
      doi            = "10.1016/j.physletb.2012.08.020",
      eprint         = "1207.7214",
      archivePrefix  = "arXiv",
      primaryClass   = "hep-ex",
      reportNumber   = "CERN-PH-EP-2012-218",
      SLACcitation   = "%%CITATION = ARXIV:1207.7214;%%"
}

@article{Chatrchyan:2012xdj,
      author         = "Chatrchyan, S. and others",
      title          = "{Observation of a new boson at a mass of 125 GeV with the
                        CMS experiment at the LHC}",
      collaboration  = "CMS",
      journal        = "Phys. Lett.",
      volume         = "B716",
      year           = "2012",
      pages          = "30-61",
      doi            = "10.1016/j.physletb.2012.08.021",
      eprint         = "1207.7235",
      archivePrefix  = "arXiv",
      primaryClass   = "hep-ex",
      reportNumber   = "CMS-HIG-12-028, CERN-PH-EP-2012-220",
      SLACcitation   = "%%CITATION = ARXIV:1207.7235;%%"
}

@article{Andreev:2018ayy,
    author = "Andreev, V. and others",
    collaboration = "ACME",
    doi = "10.1038/s41586-018-0599-8",
    journal = "Nature",
    number = "7727",
    pages = "355--360",
    title = "{Improved limit on the electric dipole moment of the electron}",
    volume = "562",
    year = "2018"
}

@article{Atlas:2019qfx,
      author        = "Dainese, Andrea and Mangano, Michelangelo and Meyer,
                       Andreas B and Nisati, Aleandro and Salam, Gavin and
                       Vesterinen, Mika Anton",
      title         = "{Report on the Physics at the HL-LHC, and Perspectives for
                       the HE-LHC}",
      address       = "Geneva, Switzerland",
      number        = "CERN-2019-007",
      year          = "2019",
      url           = "https://cds.cern.ch/record/2703572",
      journal       = "10.23731/CYRM-2019-007",
      doi           = "10.23731/CYRM-2019-007"
}

@article{Fuchs:2020uoc,
    author = "Fuchs, Elina and Losada, Marta and Nir, Yosef and Viernik, Yehonatan",
    title = "{$CP$ violation from $\tau$, $t$ and $b$ dimension-6 Yukawa couplings - interplay of baryogenesis, EDM and Higgs physics}",
    eprint = "2003.00099",
    archivePrefix = "arXiv",
    primaryClass = "hep-ph",
    reportNumber = "FERMILAB-PUB-20-072-T, EFI-20-3",
    doi = "10.1007/JHEP05(2020)056",
    journal = "JHEP",
    volume = "05",
    pages = "056",
    year = "2020"
}

@article{Kobakhidze:2016mfx,
    author = "Kobakhidze, Archil and Liu, Ning and Wu, Lei and Yue, Jason",
    title = "{Implications of CP-violating Top-Higgs Couplings at LHC and Higgs Factories}",
    eprint = "1610.06676",
    archivePrefix = "arXiv",
    primaryClass = "hep-ph",
    doi = "10.1103/PhysRevD.95.015016",
    journal = "Phys. Rev. D",
    volume = "95",
    number = "1",
    pages = "015016",
    year = "2017"
}

@article{Khachatryan:2014kca,
    author = "Khachatryan, Vardan and others",
    collaboration = "CMS",
    title = "{Constraints on the spin-parity and anomalous HVV couplings of the Higgs boson in proton collisions at 7 and 8 TeV}",
    eprint = "1411.3441",
    archivePrefix = "arXiv",
    primaryClass = "hep-ex",
    reportNumber = "CMS-HIG-14-018, CERN-PH-EP-2014-265",
    doi = "10.1103/PhysRevD.92.012004",
    journal = "Phys. Rev. D",
    volume = "92",
    number = "1",
    pages = "012004",
    year = "2015"
}

@article{Aad:2015mxa,
    author = "Aad, Georges and others",
    collaboration = "ATLAS",
    title = "{Study of the spin and parity of the Higgs boson in diboson decays with the ATLAS detector}",
    eprint = "1506.05669",
    archivePrefix = "arXiv",
    primaryClass = "hep-ex",
    reportNumber = "CERN-PH-EP-2015-114",
    doi = "10.1140/epjc/s10052-015-3685-1",
    journal = "Eur. Phys. J. C",
    volume = "75",
    number = "10",
    pages = "476",
    year = "2015",
    note = "[Erratum: Eur.Phys.J.C 76, 152 (2016)]"
}

@article{Abel:2020gbr,
    author = "Abel, C. and others",
    collaboration = "nEDM",
    title = "{Measurement of the permanent electric dipole moment of the neutron}",
    eprint = "2001.11966",
    archivePrefix = "arXiv",
    primaryClass = "hep-ex",
    doi = "10.1103/PhysRevLett.124.081803",
    journal = "Phys. Rev. Lett.",
    volume = "124",
    number = "8",
    pages = "081803",
    year = "2020"
}

@article{Aad:2020mnm,
    author = "Aad, Georges and others",
    collaboration = "ATLAS",
    title = "{Test of CP invariance in vector-boson fusion production of the Higgs boson in the $H\rightarrow\tau\tau$ channel in proton$-$proton collisions at $\sqrt{s}$ = 13 TeV with the ATLAS detector}",
    eprint = "2002.05315",
    archivePrefix = "arXiv",
    primaryClass = "hep-ex",
    reportNumber = "CERN-EP-2020-009",
    doi = "10.1016/j.physletb.2020.135426",
    journal = "Phys. Lett. B",
    volume = "805",
    pages = "135426",
    year = "2020"
}

@article{Sirunyan:2019nbs,
    author = "Sirunyan, Albert M and others",
    collaboration = "CMS",
    title = "{Constraints on anomalous $HVV$ couplings from the production of Higgs bosons decaying to $\tau$ lepton pairs}",
    eprint = "1903.06973",
    archivePrefix = "arXiv",
    primaryClass = "hep-ex",
    reportNumber = "CMS-HIG-17-034, CERN-EP-2019-029",
    doi = "10.1103/PhysRevD.100.112002",
    journal = "Phys. Rev. D",
    volume = "100",
    number = "11",
    pages = "112002",
    year = "2019"
}

@article{Sirunyan:2017tqd,
    author = "Sirunyan, Albert M and others",
    collaboration = "CMS",
    title = "{Constraints on anomalous Higgs boson couplings using production and decay information in the four-lepton final state}",
    eprint = "1707.00541",
    archivePrefix = "arXiv",
    primaryClass = "hep-ex",
    reportNumber = "CMS-HIG-17-011, CERN-EP-2017-143",
    doi = "10.1016/j.physletb.2017.10.021",
    journal = "Phys. Lett. B",
    volume = "775",
    pages = "1--24",
    year = "2017"
}

@article{Panico:2018hal,
    author = "Panico, Giuliano and Pomarol, Alex and Riembau, Marc",
    title = "{EFT approach to the electron Electric Dipole Moment at the two-loop level}",
    eprint = "1810.09413",
    archivePrefix = "arXiv",
    primaryClass = "hep-ph",
    reportNumber = "DESY-18-185",
    doi = "10.1007/JHEP04(2019)090",
    journal = "JHEP",
    volume = "04",
    pages = "090",
    year = "2019"
}

@article{Hou:2018uvr,
    author = "Hou, Wei-Shu and Kohda, Masaya and Modak, Tanmoy",
    title = "{Probing for extra top Yukawa couplings in light of $t\bar th(125)$ observation}",
    eprint = "1806.06018",
    archivePrefix = "arXiv",
    primaryClass = "hep-ph",
    doi = "10.1103/PhysRevD.98.075007",
    journal = "Phys. Rev. D",
    volume = "98",
    number = "7",
    pages = "075007",
    year = "2018"
}

@article{Cirigliano:2016nyn,
    author = "Cirigliano, V. and Dekens, W. and de Vries, J. and Mereghetti, E.",
    title = "{Constraining the top-Higgs sector of the Standard Model Effective Field Theory}",
    eprint = "1605.04311",
    archivePrefix = "arXiv",
    primaryClass = "hep-ph",
    reportNumber = "NIKHEF-2016-017, LA-UR-16-23433, NIKHEF 2016-017",
    doi = "10.1103/PhysRevD.94.034031",
    journal = "Phys. Rev. D",
    volume = "94",
    number = "3",
    pages = "034031",
    year = "2016"
}

@article{Chien:2015xha,
    author = "Chien, Y.T. and Cirigliano, V. and Dekens, W. and de Vries, J. and Mereghetti, E.",
    title = "{Direct and indirect constraints on CP-violating Higgs-quark and Higgs-gluon interactions}",
    eprint = "1510.00725",
    archivePrefix = "arXiv",
    primaryClass = "hep-ph",
    reportNumber = "LA-UR-15-27548",
    doi = "10.1007/JHEP02(2016)011",
    journal = "JHEP",
    volume = "02",
    pages = "011",
    year = "2016"
}

@article{deVries:2017ncy,
    author = "de Vries, Jordy and Postma, Marieke and van de Vis, Jorinde and White, Graham",
    title = "{Electroweak Baryogenesis and the Standard Model Effective Field Theory}",
    eprint = "1710.04061",
    archivePrefix = "arXiv",
    primaryClass = "hep-ph",
    reportNumber = "Nikhef-2017-044",
    doi = "10.1007/JHEP01(2018)089",
    journal = "JHEP",
    volume = "01",
    pages = "089",
    year = "2018"
}

@article{deVries:2018tgs,
    author = "De Vries, Jordy and Postma, Marieke and van de Vis, Jorinde",
    title = "{The role of leptons in electroweak baryogenesis}",
    eprint = "1811.11104",
    archivePrefix = "arXiv",
    primaryClass = "hep-ph",
    reportNumber = "Nikhef-2018-056",
    doi = "10.1007/JHEP04(2019)024",
    journal = "JHEP",
    volume = "04",
    pages = "024",
    year = "2019"
}

@article{Buckley:2015vsa,
    author = "Buckley, Matthew R. and Goncalves, Dorival",
    title = "{Boosting the Direct CP Measurement of the Higgs-Top Coupling}",
    eprint = "1507.07926",
    archivePrefix = "arXiv",
    primaryClass = "hep-ph",
    reportNumber = "IPPP-15-49, DCPT-15-98",
    doi = "10.1103/PhysRevLett.116.091801",
    journal = "Phys. Rev. Lett.",
    volume = "116",
    number = "9",
    pages = "091801",
    year = "2016"
}

@article{Sirunyan:2019twz,
    author = "Sirunyan, Albert M and others",
    collaboration = "CMS",
    title = "{Measurements of the Higgs boson width and anomalous $HVV$ couplings from on-shell and off-shell production in the four-lepton final state}",
    eprint = "1901.00174",
    archivePrefix = "arXiv",
    primaryClass = "hep-ex",
    reportNumber = "CMS-HIG-18-002, CERN-EP-2018-329",
    doi = "10.1103/PhysRevD.99.112003",
    journal = "Phys. Rev. D",
    volume = "99",
    number = "11",
    pages = "112003",
    year = "2019"
}

@article{Gritsan:2016hjl,
    author = "Gritsan, Andrei V. and Röntsch, Raoul and Schulze, Markus and Xiao, Meng",
    title = "{Constraining anomalous Higgs boson couplings to the heavy flavor fermions using matrix element techniques}",
    eprint = "1606.03107",
    archivePrefix = "arXiv",
    primaryClass = "hep-ph",
    reportNumber = "TTP16-020, CERN-TH-2016-135",
    doi = "10.1103/PhysRevD.94.055023",
    journal = "Phys. Rev. D",
    volume = "94",
    number = "5",
    pages = "055023",
    year = "2016"
}

@article{Faroughy:2019ird,
    author = "Faroughy, Darius A. and Kamenik, Jernej F. and Ko\v{s}nik, Nejc and Smolkovi\v{c}, Aleks",
    title = "{Probing the $CP$ nature of the top quark Yukawa at hadron colliders}",
    eprint = "1909.00007",
    archivePrefix = "arXiv",
    primaryClass = "hep-ph",
    doi = "10.1007/JHEP02(2020)085",
    journal = "JHEP",
    volume = "02",
    pages = "085",
    year = "2020"
}

@article{Bortolato:2020zcg,
    author = "Bortolato, Bla\v{z} and Kamenik, Jernej F. and Ko\v{s}nik, Nejc and Smolkovi\v{c}, Aleks",
    title = "{Optimized probes of $CP$ -odd effects in the $t \bar t h$ process at hadron colliders}",
    eprint = "2006.13110",
    archivePrefix = "arXiv",
    primaryClass = "hep-ph",
    doi = "10.1016/j.nuclphysb.2021.115328",
    journal = "Nucl. Phys. B",
    volume = "964",
    pages = "115328",
    year = "2021"
}

@article{Brehmer:2018hga,
    author = "Brehmer, Johann and Louppe, Gilles and Pavez, Juan and Cranmer, Kyle",
    title = "{Mining gold from implicit models to improve likelihood-free inference}",
    eprint = "1805.12244",
    archivePrefix = "arXiv",
    primaryClass = "stat.ML",
    doi = "10.1073/pnas.1915980117",
    journal = "Proc. Nat. Acad. Sci.",
    volume = "117",
    number = "10",
    pages = "5242--5249",
    year = "2020"
}

@article{Brehmer:2018kdj,
    author = "Brehmer, Johann and Cranmer, Kyle and Louppe, Gilles and Pavez, Juan",
    title = "{Constraining Effective Field Theories with Machine Learning}",
    eprint = "1805.00013",
    archivePrefix = "arXiv",
    primaryClass = "hep-ph",
    doi = "10.1103/PhysRevLett.121.111801",
    journal = "Phys. Rev. Lett.",
    volume = "121",
    number = "11",
    pages = "111801",
    year = "2018"
}

@article{Brehmer:2018eca,
    author = "Brehmer, Johann and Cranmer, Kyle and Louppe, Gilles and Pavez, Juan",
    title = "{A Guide to Constraining Effective Field Theories with Machine Learning}",
    eprint = "1805.00020",
    archivePrefix = "arXiv",
    primaryClass = "hep-ph",
    doi = "10.1103/PhysRevD.98.052004",
    journal = "Phys. Rev. D",
    volume = "98",
    number = "5",
    pages = "052004",
    year = "2018"
}

@article{Stoye:2018ovl,
    author = "Stoye, Markus and Brehmer, Johann and Louppe, Gilles and Pavez, Juan and Cranmer, Kyle",
    title = "{Likelihood-free inference with an improved cross-entropy estimator}",
    eprint = "1808.00973",
    archivePrefix = "arXiv",
    primaryClass = "stat.ML",
    month = "8",
    year = "2018"
}

@article{Brehmer:2019bvj,
    author = "Brehmer, Johann and Cranmer, Kyle and Espejo, Irina and Kling, Felix and Louppe, Gilles and Pavez, Juan",
    title = "{Effective LHC measurements with matrix elements and machine learning}",
    eprint = "1906.01578",
    archivePrefix = "arXiv",
    primaryClass = "hep-ph",
    doi = "10.1088/1742-6596/1525/1/012022",
    journal = "J. Phys. Conf. Ser.",
    volume = "1525",
    number = "1",
    pages = "012022",
    year = "2020"
}

@article{Brehmer:2019xox,
    author = "Brehmer, Johann and Kling, Felix and Espejo, Irina and Cranmer, Kyle",
    title = "{MadMiner: Machine learning-based inference for particle physics}",
    eprint = "1907.10621",
    archivePrefix = "arXiv",
    primaryClass = "hep-ph",
    doi = "10.1007/s41781-020-0035-2",
    journal = "Comput. Softw. Big Sci.",
    volume = "4",
    number = "1",
    pages = "3",
    year = "2020"
}

@article{Bahl:2020wee,
    author = "Bahl, Henning and Bechtle, Philip and Heinemeyer, Sven and Katzy, Judith and Klingl, Tobias and Peters, Krisztian and Saimpert, Matthias and Stefaniak, Tim and Weiglein, Georg",
    title = "{Indirect $\mathcal{CP}$ probes of the Higgs-top-quark interaction: current LHC constraints and future opportunities}",
    eprint = "2007.08542",
    archivePrefix = "arXiv",
    primaryClass = "hep-ph",
    doi = "10.1007/JHEP11(2020)127",
    journal = "JHEP",
    volume = "11",
    pages = "127",
    year = "2020"
}

@article{CMS:2020rpr,
    collaboration = "CMS",
    title = "{Analysis of the CP structure of the Yukawa coupling between the Higgs boson and $\tau$ leptons in proton-proton collisions at $\sqrt{s}=13~\mathrm{TeV}$}",
    reportNumber = "CMS-PAS-HIG-20-006",
    month = "8",
    year = "2020"
}

@article{CMS:2020dkv,
    collaboration = "CMS",
    title = "{Constraints on anomalous Higgs boson couplings to vector bosons and fermions in production and decay in the $H\to4\ell$ channel}",
    reportNumber = "CMS-PAS-HIG-19-009",
    month = "8",
    year = "2020"
}

@article{Aad:2016nal,
    author = "Aad, Georges and others",
    collaboration = "ATLAS",
    title = "{Test of CP Invariance in vector-boson fusion production of the Higgs boson using the Optimal Observable method in the ditau decay channel with the ATLAS detector}",
    eprint = "1602.04516",
    archivePrefix = "arXiv",
    primaryClass = "hep-ex",
    reportNumber = "CERN-EP-2016-002",
    doi = "10.1140/epjc/s10052-016-4499-5",
    journal = "Eur. Phys. J. C",
    volume = "76",
    number = "12",
    pages = "658",
    year = "2016"
}

@techreport{ATL-PHYS-PUB-2015-047,
      title         = "{A morphing technique for signal modelling in a
                       multidimensional space of coupling parameters}",
      institution   = "CERN",
      address       = "Geneva",
      number        = "ATL-PHYS-PUB-2015-047",
      month         = "Nov",
      year          = "2015",
      reportNumber  = "ATL-PHYS-PUB-2015-047",
      url           = "http://cds.cern.ch/record/2066980",
}

@article{CMS:2021nnc,
    author = "Sirunyan, Albert M and others",
    collaboration = "CMS",
    title = "{Constraints on anomalous Higgs boson couplings to vector bosons and fermions in its production and decay using the four-lepton final state}",
    eprint = "2104.12152",
    archivePrefix = "arXiv",
    primaryClass = "hep-ex",
    reportNumber = "CMS-HIG-19-009, CERN-EP-2021-054",
    month = "4",
    year = "2021"
}

@article{Martini:2021uey,
    author = "Martini, Till and Pan, Ren-Qi and Schulze, Markus and Xiao, Meng",
    title = "{Probing the CP structure of the top quark Yukawa coupling: Loop sensitivity versus on-shell sensitivity}",
    eprint = "2104.04277",
    archivePrefix = "arXiv",
    primaryClass = "hep-ph",
    reportNumber = "HU-EP-21/06",
    doi = "10.1103/PhysRevD.104.055045",
    journal = "Phys. Rev. D",
    volume = "104",
    number = "5",
    pages = "055045",
    year = "2021"
}

@article{Bolognesi:2012mm,
    author = "Bolognesi, Sara and Gao, Yanyan and Gritsan, Andrei V. and Melnikov, Kirill and Schulze, Markus and Tran, Nhan V. and Whitbeck, Andrew",
    title = "{On the spin and parity of a single-produced resonance at the LHC}",
    eprint = "1208.4018",
    archivePrefix = "arXiv",
    primaryClass = "hep-ph",
    reportNumber = "ANL-HEP-PR-12-62, FERMILAB-PUB-12-475-PPD",
    doi = "10.1103/PhysRevD.86.095031",
    journal = "Phys. Rev. D",
    volume = "86",
    pages = "095031",
    year = "2012"
}

@article{Avery:2012um,
    author = "Avery, Paul and others",
    title = "{Precision studies of the Higgs boson decay channel H\textrightarrow{}ZZ\textrightarrow{}4\ensuremath{\ell} with MEKD}",
    eprint = "1210.0896",
    archivePrefix = "arXiv",
    primaryClass = "hep-ph",
    reportNumber = "CERN-PH-TH-2012-251",
    doi = "10.1103/PhysRevD.87.055006",
    journal = "Phys. Rev. D",
    volume = "87",
    number = "5",
    pages = "055006",
    year = "2013"
}

@article{Artoisenet:2013vfa,
    author = "Artoisenet, Pierre and de Aquino, Priscila and Maltoni, Fabio and Mattelaer, Olivier",
    title = "{Unravelling $t\overline{t}h$ via the Matrix Element Method}",
    eprint = "1304.6414",
    archivePrefix = "arXiv",
    primaryClass = "hep-ph",
    reportNumber = "NIKHEF-2013-011, CP3-13-14",
    doi = "10.1103/PhysRevLett.111.091802",
    journal = "Phys. Rev. Lett.",
    volume = "111",
    number = "9",
    pages = "091802",
    year = "2013"
}

@inproceedings{Gainer:2013iya,
    author = "Gainer, James S. and Lykken, Joseph and Matchev, Konstantin T. and Mrenna, Stephen and Park, Myeonghun",
    title = "{The Matrix Element Method: Past, Present, and Future}",
    booktitle = "{Community Summer Study 2013}: {Snowmass on the Mississippi}",
    eprint = "1307.3546",
    archivePrefix = "arXiv",
    primaryClass = "hep-ph",
    reportNumber = "CERN-PH-TH-2013-165, FERMILAB-FN-0963-CD-T",
    month = "7",
    year = "2013"
}

@article{Chang:2014rfa,
    author = "Chang, Jung and Cheung, Kingman and Lee, Jae Sik and Lu, Chih-Ting",
    title = "{Probing the Top-Yukawa Coupling in Associated Higgs production with a Single Top Quark}",
    eprint = "1403.2053",
    archivePrefix = "arXiv",
    primaryClass = "hep-ph",
    reportNumber = "CNU-HEP-14-01",
    doi = "10.1007/JHEP05(2014)062",
    journal = "JHEP",
    volume = "05",
    pages = "062",
    year = "2014"
}

@article{Yue:2014tya,
    author = "Yue, Jason",
    title = "{Enhanced $thj$ signal at the LHC with $h\rightarrow \gamma\gamma$ decay and $\mathcal{CP}$-violating top-Higgs coupling}",
    eprint = "1410.2701",
    archivePrefix = "arXiv",
    primaryClass = "hep-ph",
    doi = "10.1016/j.physletb.2015.03.044",
    journal = "Phys. Lett. B",
    volume = "744",
    pages = "131--136",
    year = "2015"
}

@article{Brehmer:2016nyr,
    author = "Brehmer, Johann and Cranmer, Kyle and Kling, Felix and Plehn, Tilman",
    title = "{Better Higgs boson measurements through information geometry}",
    eprint = "1612.05261",
    archivePrefix = "arXiv",
    primaryClass = "hep-ph",
    doi = "10.1103/PhysRevD.95.073002",
    journal = "Phys. Rev. D",
    volume = "95",
    number = "7",
    pages = "073002",
    year = "2017"
}

@article{Brehmer:2017lrt,
    author = "Brehmer, Johann and Kling, Felix and Plehn, Tilman and Tait, Tim M. P.",
    title = "{Better Higgs-CP Tests Through Information Geometry}",
    eprint = "1712.02350",
    archivePrefix = "arXiv",
    primaryClass = "hep-ph",
    reportNumber = "UCI-HEP-TR-2017-14",
    doi = "10.1103/PhysRevD.97.095017",
    journal = "Phys. Rev. D",
    volume = "97",
    number = "9",
    pages = "095017",
    year = "2018"
}

@article{Rao:1945,
    author = "Rao, C. R.",
    title = "{Information and the accuracy attainable in the estimation of statistical parameters}",
    journal = "Bull. Calcutta Math. Soc.",
    volume = "37",
    pages = "81",
    year = "1945"
}

@book{Cramer:1946,
    author = "Cramér, H.",
    title = "{Mathematical Methods of Statistics}",
    publisher = "Princeton University Press",
    ISBN = "0691080046.",
    year = "1946"
}

@article{ATLAS:2020pvn,
    collaboration = "ATLAS",
    title = "{Measurement of the properties of Higgs boson production at $\sqrt{s}$=13 TeV in the $H\to \gamma\gamma$ channel using 139 fb$^{-1}$ of $pp$ collision data with the ATLAS experiment}",
    reportNumber = "ATLAS-CONF-2020-026",
    month = "8",
    year = "2020"
}

@article{Goncalves:2021dcu,
    author = "Gon\c{c}alves, Dorival and Kim, Jeong Han and Kong, Kyoungchul and Wu, Yongcheng",
    title = "{Direct Higgs-top CP-phase measurement with $t\bar{t}h$ at the 14 TeV LHC and 100 TeV FCC}",
    eprint = "2108.01083",
    archivePrefix = "arXiv",
    primaryClass = "hep-ph",
    month = "8",
    year = "2021"
}

@article{lakshminarayanan2017simple,
      title={Simple and Scalable Predictive Uncertainty Estimation using Deep Ensembles},
      author={Balaji Lakshminarayanan and Alexander Pritzel and Charles Blundell},
      year={2017},
      eprint={1612.01474},
      archivePrefix={arXiv},
      primaryClass={stat.ML}
}

@article{CMS:2021sdq,
    author = "Tumasyan, Armen and others",
    collaboration = "CMS",
    title = "{Analysis of the CP structure of the Yukawa coupling between the Higgs boson and $\tau$ leptons in proton-proton collisions at $\sqrt{s}$ = 13 TeV}",
    eprint = "2110.04836",
    archivePrefix = "arXiv",
    primaryClass = "hep-ex",
    reportNumber = "CMS-HIG-20-006, CERN-EP-2021-189",
    month = "10",
    year = "2021"
}

@article{Barman:2021yfh,
    author = "Barman, Rahool Kumar and Gon\c{c}alves, Dorival and Kling, Felix",
    title = "{Machine Learning the Higgs-Top CP Phase}",
    eprint = "2110.07635",
    archivePrefix = "arXiv",
    primaryClass = "hep-ph",
    reportNumber = "DESY 21-161",
    month = "10",
    year = "2021"
}

@article{Ren:2019xhp,
    author = "Ren, Jie and Wu, Lei and Yang, Jin Min",
    title = "{Unveiling CP property of top-Higgs coupling with graph neural networks at the LHC}",
    eprint = "1901.05627",
    archivePrefix = "arXiv",
    primaryClass = "hep-ph",
    doi = "10.1016/j.physletb.2020.135198",
    journal = "Phys. Lett. B",
    volume = "802",
    pages = "135198",
    year = "2020"
}

@article{Agrawal:2012ga,
    author = "Agrawal, Pankaj and Mitra, Subhadip and Shivaji, Ambresh",
    title = "{Effect of Anomalous Couplings on the Associated Production of a Single Top Quark and a Higgs Boson at the LHC}",
    eprint = "1211.4362",
    archivePrefix = "arXiv",
    primaryClass = "hep-ph",
    reportNumber = "LPT-ORSAY-12-112",
    doi = "10.1007/JHEP12(2013)077",
    journal = "JHEP",
    volume = "12",
    pages = "077",
    year = "2013"
}

@article{Mileo:2016mxg,
    author = "Mileo, Nicolas and Kiers, Ken and Szynkman, Alejandro and Crane, Daniel and Gegner, Ethan",
    title = "{Pseudoscalar top-Higgs coupling: exploration of CP-odd observables to resolve the sign ambiguity}",
    eprint = "1603.03632",
    archivePrefix = "arXiv",
    primaryClass = "hep-ph",
    doi = "10.1007/JHEP07(2016)056",
    journal = "JHEP",
    volume = "07",
    pages = "056",
    year = "2016"
}

@article{Martini:2015fsa,
    author = "Martini, Till and Uwer, Peter",
    title = "{Extending the Matrix Element Method beyond the Born approximation: Calculating event weights at next-to-leading order accuracy}",
    eprint = "1506.08798",
    archivePrefix = "arXiv",
    primaryClass = "hep-ph",
    reportNumber = "HU-EP-15-28",
    doi = "10.1007/JHEP09(2015)083",
    journal = "JHEP",
    volume = "09",
    pages = "083",
    year = "2015"
}

@article{Martini:2017ydu,
    author = "Martini, Till and Uwer, Peter",
    title = "{The Matrix Element Method at next-to-leading order QCD for hadronic collisions: Single top-quark production at the LHC as an example application}",
    eprint = "1712.04527",
    archivePrefix = "arXiv",
    primaryClass = "hep-ph",
    reportNumber = "HU-EP-17-28",
    doi = "10.1007/JHEP05(2018)141",
    journal = "JHEP",
    volume = "05",
    pages = "141",
    year = "2018"
}

@article{Kraus:2019qoq,
    author = "Kraus, Manfred and Martini, Till and Uwer, Peter",
    title = "{Matrix Element Method at NLO for {(anti-)$\mathbf{k_t}$}-jet algorithms}",
    eprint = "1901.08008",
    archivePrefix = "arXiv",
    primaryClass = "hep-ph",
    reportNumber = "HU-EP-18/35",
    doi = "10.1103/PhysRevD.100.076010",
    journal = "Phys. Rev. D",
    volume = "100",
    number = "7",
    pages = "076010",
    year = "2019"
}

@article{D0:2004rvt,
    author = "Abazov, V. M. and others",
    collaboration = "D0",
    title = "{A precision measurement of the mass of the top quark}",
    eprint = "hep-ex/0406031",
    archivePrefix = "arXiv",
    reportNumber = "FERMILAB-PUB-04-083-E",
    doi = "10.1038/nature02589",
    journal = "Nature",
    volume = "429",
    pages = "638--642",
    year = "2004"
}

@article{Gao:2010qx,
    author = "Gao, Yanyan and Gritsan, Andrei V. and Guo, Zijin and Melnikov, Kirill and Schulze, Markus and Tran, Nhan V.",
    title = "{Spin Determination of Single-Produced Resonances at Hadron Colliders}",
    eprint = "1001.3396",
    archivePrefix = "arXiv",
    primaryClass = "hep-ph",
    reportNumber = "FERMILAB-PUB-10-011-E",
    doi = "10.1103/PhysRevD.81.075022",
    journal = "Phys. Rev. D",
    volume = "81",
    pages = "075022",
    year = "2010"
}

@article{Alwall:2010cq,
    author = "Alwall, J. and Freitas, A. and Mattelaer, O.",
    title = "{The Matrix Element Method and QCD Radiation}",
    eprint = "1010.2263",
    archivePrefix = "arXiv",
    primaryClass = "hep-ph",
    reportNumber = "CP3-10-35",
    doi = "10.1103/PhysRevD.83.074010",
    journal = "Phys. Rev. D",
    volume = "83",
    pages = "074010",
    year = "2011"
}

@article{Andersen:2012kn,
    author = "Andersen, Jeppe R. and Englert, Christoph and Spannowsky, Michael",
    title = "{Extracting precise Higgs couplings by using the matrix element method}",
    eprint = "1211.3011",
    archivePrefix = "arXiv",
    primaryClass = "hep-ph",
    reportNumber = "IPPP-12-86, DCPT-12-172",
    doi = "10.1103/PhysRevD.87.015019",
    journal = "Phys. Rev. D",
    volume = "87",
    number = "1",
    pages = "015019",
    year = "2013"
}

@article{Campbell:2013hz,
    author = "Campbell, John M. and Ellis, R. Keith and Giele, Walter T. and Williams, Ciaran",
    title = "{Finding the Higgs boson in decays to $Z \gamma$ using the matrix element method at Next-to-Leading Order}",
    eprint = "1301.7086",
    archivePrefix = "arXiv",
    primaryClass = "hep-ph",
    reportNumber = "FERMILAB-PUB-13-024-T",
    doi = "10.1103/PhysRevD.87.073005",
    journal = "Phys. Rev. D",
    volume = "87",
    number = "7",
    pages = "073005",
    year = "2013"
}

@article{Schouten:2014yza,
    author = "Schouten, Doug and DeAbreu, Adam and Stelzer, Bernd",
    title = "{Accelerated Matrix Element Method with Parallel Computing}",
    eprint = "1407.7595",
    archivePrefix = "arXiv",
    primaryClass = "physics.comp-ph",
    doi = "10.1016/j.cpc.2015.02.020",
    journal = "Comput. Phys. Commun.",
    volume = "192",
    pages = "54--59",
    year = "2015"
}

@article{ATLAS:2021pkb,
    author = "Aad, Georges and others",
    collaboration = "ATLAS",
    title = "{Constraints on Higgs boson properties using $WW^{*}(\rightarrow e\nu\mu\nu) jj$ production in 36.1 fb$^{-1}$ of $\sqrt{s}$=13 TeV $pp$ collisions with the ATLAS detector}",
    eprint = "2109.13808",
    archivePrefix = "arXiv",
    primaryClass = "hep-ex",
    reportNumber = "CERN-EP-2021-096",
    month = "9",
    year = "2021"
}
